%% file: paper.tex
\documentclass[runningheads]{llncs}
\usepackage{graphicx}
\usepackage{booktabs}
\usepackage{multirow}
\usepackage{amsmath} 
\usepackage{verbatim}
\usepackage{float}
\usepackage{subfigure}
\usepackage[noend]{algpseudocode}
\usepackage{amsfonts}
\usepackage{xcolor}
\usepackage{algorithm}
\usepackage[noend]{ }

\usepackage{textcomp}

\DeclareMathOperator*{\argmax}{arg\,max}

\allowdisplaybreaks

\begin{document}
\title{Blocking Adversarial Influence in Social Networks}
%
%

\author{
	Feiran Jia\inst{1}
	\and
	Kai Zhou\inst{1}
	\and
	Charles Kamhoua\inst{2}
	\and
	Yevgeniy Vorobeychik\inst{1}
}
\authorrunning{F. Jia et al.}
%
\institute{Department of Computer Science and Engineering, Washington University in St.\ Louis,  St. Louis, MO 63130, USA \and
Army Research Laboratory, 2800 Powder Mill Rd, Adelphi, MD 20783, USA
\email{\{feiran.jia, zhoukai, yvorobeychik\}@wustl.edu}\\
\email{charles.a.kamhoua.civ@mail.mil}}

\maketitle              
\begin{abstract}
While social networks are widely used as a media for information diffusion, attackers can also strategically employ analytical tools, such as influence maximization, to maximize the spread of adversarial content through the networks. 
We investigate the problem of limiting the diffusion of negative
information by blocking nodes and edges in the network. We formulate the
interaction between the defender and the attacker as a Stackelberg
game where the defender first chooses a set of nodes to block and then
the attacker selects a set of seeds to spread negative information
from. 
This yields an extremely complex bi-level optimization problem,
particularly since even the standard influence measures are difficult
to compute.
Our approach is to approximate the attacker's problem as the maximum
node domination problem.
To solve this problem, we first develop a method based on integer
programming combined with constraint generation.
Next, to improve scalability, we develop an approximate solution
method that represents the attacker's problem as an integer program,
and then combines relaxation with duality to yield an upper bound on
the defender's objective that can be computed using mixed integer
linear programming.
Finally, we propose an even more scalable heuristic method that prunes
nodes from the consideration set based on their degree.
Extensive experiments demonstrate the efficacy of our approaches.
\end{abstract}


\input{./sections/intro}

\input{./sections/prob}

\input{./sections/solution}

\input{./sections/extensions}

\input{./sections/exp}

\section{Conclusion}

In this paper, we investigate the problem of blocking adversarial
information in social networks, where a network defender aims to limit
the spread of misinformation by blocking nodes in the network. We
model the problem as a Stackelberg game and seek the optimal strategy
for the defender. The main challenge is to find the best response for
the attacker, which involves solving the influence maximization
problem. Our approach is to approximate the attacker's influence
maximization as the maximum node domination problem, which can be
expressed as an integer program. This enables us to develop a
constraint generation approach for the defender's problem.  Further, by utilizing linear program relaxation and its duality, we reformulate the defender's problem as a mixed-integer linear program, which can be solved efficiently.  We further develop a heuristic pruning algorithm to deal with large networks efficiently, as well as a constraint generation algorithm to compute the exact solution iteratively. 

We test our defense approaches against several attacks on synthetic graphs and real-world networks and compare them with various state-of-the-art defense baselines. The experiment results show that our proposed defense approaches can effectively limit the spread of misinformation in an adversarial environment, outperforming all other baselines.

\subsection*{Acknowledgment}
This research was partially supported by the NSF (IIS-1903207 and CAREER Grant IIS-1905558) and ARO MURI (W911NF1810208).

\bibliographystyle{splncs04}
\small
\bibliography{Bibliography-File.bib}
%




\end{document}

%% file: sections/intro.tex
\section{Introduction}

The problem of diffusion over social networks has received considerable prior attention in the literature, both from the perspective of promoting diffusion (the so-called \emph{influence maximization} problem) as well as in preventing its spread (the \emph{influence blocking} problem).
The influence maximization problem aims to select a subset of nodes on a network to maximize the overall spread of influence, such as adoption of a product or an opinion~\cite{Domingos:2001:MNV:502512.502525,Kempe:2003:MSI:956750.956769}.
Influence blocking presumes that a diffusion process is spreading, typically either from a set of known nodes, or from nodes selected according to some known distribution, with the goal of blocking its path through either a select set of nodes or edges~\cite{KimuraSM08,Kimura:2008:MSC:1620163.1620255,Kimura:2009:BLM:1514888.1514892,Khalil:2014:SDO:2623330.2623704,Wang2013NegativeIM,Yao:2015:TSI:2740908.2742767}.

In many settings, influence maximizers are malicious parties, and our goal is to limit their overall influence.
For example, in cybersecurity, influence maximization may correspond to the spread of malware on the network, while in criminology we may be concerned about the spread of criminal influence (such as promoting membership in gangs or terrorist organizations).
It is natural in these settings to consider the problem of \emph{adversarial influence blocking (AIB)}, where a defender can first block (inoculate) a set of nodes or edges, and the adversary subsequently unleashes an influence maximization process.
In the cybersecurity setting, we may impose use restrictions on a subset of computing devices, or even island these from the internet.

We model the resulting problem as a Stackelberg security game in which the defender first chooses (deterministically) which subset of nodes to block, and the attacker then selects a subset of seed nodes to begin network diffusion.
The adversary's goal is to maximize overall influence, whereas the defender aims to minimize it.
Note that this problem is significantly more difficult than the traditional influence blocking problem, since we are now allowing the choice of seeds to be adversarial, and to condition on the nodes we choose to block.
Despite the extensive prior research on both influence maximization and influence blocking problems and their many variants, however, no general effective solution exists for the adversarial influence blocking problem.

The AIB problem is an extremely challenging bi-level optimization problem for a host of reasons.
First, even computing influence for general influence measures is difficult~\cite{Chen:2010:SIM:1933307.1934561,Chen:2010:SIM:1835804.1835934}.
Moreover, influence maximization is hard even if we assume that we can use a black-box (e.g., simulations) to compute expected influence, and it's only a subproblem.
To address these technical challenges, we first approximate influence maximization in the lower-level problem by a \emph{maximum node domination problem}.
While this problem is still NP-Hard~\cite{Miyano:2011:MDP:2483191.2483198}, it can be solved using integer linear programming (ILP).
We make use of this, together with a constraint generation algorithm, to develop the first practical solution to AIM.
To increase scalability, we develop an approximation based on a relaxation of the attacker's ILP combined with duality, which yields a single-level mixed-integer linear program for the defender to solve.
We further improve the scalability of the resulting approach by using simple node pruning heuristics (removing a subset of nodes from consideration in the optimization problem).
Through extensive experiments, we show that our approach is more effective for computing influence blocking nodes than state of the art alternatives for a variety of influence measures, including domination, independent cascades, and linear threshold models.

\paragraph{Related Work}

\emph{Influence maximization} (IM) is a classical problem in social network analysis, which aims at identifying a set of seeds to maximize the spread of influence under an information diffusion model, such as the independent cascade (IC) and linear threshold (LT) model. It has been shown that identifying such a seed set is NP-hard and proposed a greedy algorithm with provable guarantees\cite{Kempe:2003:MSI:956750.956769}. 

On the contrary, a host of works consider the \emph{influence blocking} problem of limiting the spread of information, typically through blocking the paths of diffusion, or equivalently modifying the underlying network structure. Some of them considered removing the edges, with the goal of minimizing the averaged influence of all nodes by treating each node as a seed independently~\cite{KimuraSM08,Kimura:2008:MSC:1620163.1620255,Kimura:2009:BLM:1514888.1514892}, or minimizing the overall influence of a known set of sources
~\cite{Khalil:2014:SDO:2623330.2623704}.
Most of these works proposed heuristic algorithms, and experimentally demonstrated the efficacy under the LT or/add IC models. An exception is that the objective function under the LT model is supermudular, resulting in scalable and effective algorithms~\cite{Khalil_cuttingedge:influence,Khalil:2014:SDO:2623330.2623704}. 
There are also other works considering removing nodes from the network and proposed several heuristic approaches based on the node properties, such as out-degrees~\cite{albert2000error,Newman:2002email,Callaway_2000} and betweenness centrality~\cite{Yao:2015:TSI:2740908.2742767}. However, all these works consider a rather \textit{static} scenario, where the initial set of seeds is known and fixed, which is fundamentally different from ours.

Besides modifying the network structure, an orthogonal line of works ~\cite{Budak:2011:LSM:1963405.1963499,conf/sdm/HeSCJ12}
consider the problem of spreading positive information as the best response to limit the eventual negative influence caused by the static adversary.
Other works focus on the game-theoretic version where both the players choose to propagate their influence strategically and simultaneously~\cite{Tsai:2012:SGC:2900929.2900936,10.1093/comjnl/bxt094,6693308}.
Several following works model such a setting as games between the two sources in various application scenarios such as the defending against misinformation in elections~\cite{Wilder_Vorobeychik_2019} and protecting assets in an interdependent setting~\cite{Vorobeychik:2015:SIA:2738482.2738535}.

Our approach relies on approximating the influence of maximization as the \emph{Maximum Node Domination problem}, which we term as \textit{k-MaxVD}. 
In a graph, the set of dominated nodes of a node $i$ includes $i$ and its neighbors. 
The Node Domination Set \cite{Garey} of a node-set $U$ is then the union of all the dominated nodes of every node in $U$. The \textit{k-MaxVD} problem is then to find the set $U$ of $k$ nodes such that the size of its Node Domination Set is maximized. 
\textit{k-MaxVD} is proved to be NP-hard, and a simple greedy algorithm achieves an approximation ratio of $(1-1/e)$~\cite{Miyano:2011:MDP:2483191.2483198}.

%% file: sections/prob.tex
\section{Problem Formulation}
In this section, we formulate the \emph{adversarial influence blocking} problem as a Stackelberg game where the attacker solves the influence maximization problem after observing a network modified by the defender. To make it tractable, we approximate the attacker's problem as the maximum node domination (k-MaxVD) problem.
\paragraph{Stackelberg game model}
We consider a graph $\mathcal{G} = (V,E)$, with a set of $n$ nodes $V$ and a set of $m$ edges $E$. 
A defender selects a set of nodes $S_D \subseteq V$ to block (remove from the graph) aiming at minimizing the negative influence caused by the attacker. 
We use $\mathcal{G}(S_D)$ to denote the modified graph after nodes in $S_D$ are blocked. 
After observing $\mathcal{G}(S_D)$, an attacker selects an initial set of seeds $S_A$ to maximize the influence under a given influence diffusion model. 
Since the attacker's strategy is conditioned on the choice of $S_D$, we represent it as a function $g(S_D)$.  
The interaction between the defender and the attacker is formulated as a Stackelberg game with the defender as the leader and the attacker the follower.
To formalize, we denote the utilities of the defender and the attacker as $U_D (S_D,S_A)$ and $U_A (S_D,S_A)$, respectively. Our goal is thus to seek the Stackelberg Equilibrium (SE) of the game, which is defined as follows:
\begin{definition}
A strategy profile $(S_D^*, g^*(S_D))$ forms a Stackelberg Equilibrium of the game if it satisfies two conditions:
\begin{itemize}
	\item The defender plays a best response:
	$$ U_D(S_D^*,g^*(S_D^*)) \geq U_D(S_D,g^*(S_D)), \forall S_D.$$
	\item The attacker plays a best response to $S_D$:
	$$ U_A(S_D,g^*(S_D)) \geq U_A(S_D,g(S_D)), \forall g, S_D.$$
\end{itemize}
\end{definition}
In particular, we focus on approximating a Strong Stackelberg equilibrium (SSE), in which the attacker breaks ties (if any) in the defender's favor.

Next, we define the utilities for both players in terms of the results of adversarial influence on the network.
Specifically, the \textit{influence} of a seed set $S_A$ chosen by the attacker is the total number of influenced nodes resulting from an exogenously specified diffusion model, denoted by $\sigma (S_A|\mathcal{G}(S_D))$.
The particular game we consider is a zero-sum game in which the attacker's utility is the influence $\sigma (S_A|\mathcal{G}(S_D))$; formally,
$U_D (S_D,S_A) = - \sigma (S_A|\mathcal{G}(S_D))$ and  $U_A (S_D,S_A) = \sigma (S_A|\mathcal{G}(S_D))$.
A key concept in this model is the \emph{influence maximization problem}, $\mathsf{InfluMax}(\mathcal{G})$, which takes a graph $\mathcal{G}$ as input and outputs an optimal set of seeds; this is the \emph{attacker's problem}.
Consequently, finding the SSE of the game involves solving the following bi-level program:
\begin{align}
\label{eqn-bi-level-original}
\min_{S_D} &\quad   \sigma(S_A^*|\mathcal{G}(S_D)) \\
\text{s.t.} &\quad |S_D| \leq k_D \nonumber \\
&\quad  S_A^* = \mathsf{InfluMax}(\mathcal{G}(S_D)) \nonumber \\
&\quad \text{s.t.} \quad |S_A| \leq k_A, \nonumber
\end{align}
where $k_A$ and $k_D$ are budget constraints on $S_A$ and $S_D$, the sets of nodes the attacker can influence, and the defender can block (remove from the graph), respectively.

It is evident that the bi-level program~\eqref{eqn-bi-level-original} is quite intractable, first because common influence measures, such as using the independent cascades model, are intractable to compute, second because influence maximization is itself NP-Hard, and third because both the outer and inner optimization problems are non-convex.
Furthermore, given that there are many competing models of diffusion of influence on networks, there is even ambiguity in how to best instantiate the influence function $\sigma (S_A|\mathcal{G}(S_D))$.
For these reasons, we next propose an approximation of the influence functions that introduces considerably more structure to the problem, and that can be a proxy for many conventional influence functions in the literature.

\paragraph{Approximating the influence} Solving the previous bi-level program involves solving $\mathsf{InfluMax}(\mathcal{G}(S_D))$ given any $S_D$. However, finding the optimal seed set $S_A$ that maximizes $\sigma (S_A|\mathcal{G}(S_D))$ is NP-hard for essentially any common influence measure~\cite{Kempe:2003:MSI:956750.956769}. In fact, even mathematically formulating $\mathsf{InfluMax}(\mathcal{G}(S_D))$ is not easy -- the typical approaches treat $\mathsf{InfluMax}(\mathcal{G}(S_D))$ as a black box and identify the optimal $S_A$ through simulation. 
To make our problem more tractable, we approximate $\sigma (S_A|\mathcal{G}(S_D))$ as the cardinality of the \textit{dominated node set} with respect to $S_A$, denoted by $\mathcal{D}(S_A|\mathcal{G}(S_D))$.
Specifically, given a node $v \in V$, its dominated node set is defined as the $v$ and its neighbors in the graph, i.e., $\mathcal{D}_v = v \cup N(v)$, where $N(v)$ is the set of neighbors of $v$. 
Then the dominated node set of $S_A$ is defined as 
\begin{align*}
 \mathcal{D}(S_A|\mathcal{G}(S_D)) = \cup_{v \in S_A} \mathcal{D}_v  = \{u| \exists v \in S_A, \text{s.t.} (u,v) \in E\},
\end{align*}
and we approximate the influence function using the cardinality of this set: $\sigma (S_A|\mathcal{G}(S_D)) \approx |\mathcal{D}(S_A|\mathcal{G}(S_D))|$.
As a result, the influence maximization problem $\mathsf{InfluMax}(\mathcal{G}(S_D))$ is approximated as the maximum node domination problem, which is to find the node set $S_A$ that maximizes $\mathcal{D}(S_A|\mathcal{G}(S_D))$.
The resulting bi-level problem we aim to solve is
\begin{align}
\label{Eqn-bi-level}
\min_{S_D} &\quad   |\mathcal{D}(S_A^*|\mathcal{G}(S_D))| \\
\text{s.t.} &\quad |S_D| \leq k_D \nonumber \\
            &\quad  S_A^* = \argmax_{S_A} \quad |\mathcal{D}(S_A|\mathcal{G}(S_D))| \nonumber \\
            &\quad \text{s.t.} \quad |S_A| \leq k_A. \nonumber
\end{align}
The solution to problem~\eqref{Eqn-bi-level} then becomes the approximate solution to problem~\eqref{eqn-bi-level-original}.
We note that approximation here is not formal; rather, we use experiments below to show its effectiveness in comparison with a number of alternatives.
Moreover, node domination is itself a natural influence measure (as a generalization of a node's degree centrality).

%% file: sections/solution.tex
\section{Solution Approach}

In this section, we present several approaches for computing the
defender's optimal strategy.
To begin, we rewrite the bi-level problem as follows.
Denote the defender's strategy as a binary vector $\mathbf{x} = \{0,1\}^n$, where $x_i = 1$ means that the defender chooses to block node $v_i$ and $x_i = 0$ otherwise. 
Similarly, let $\mathbf{y} = \{0,1\}^n$ denote the attacker's strategy, where $y_i = 1$ means that the attacker selects $v_i$ as a seed and $y_i = 0$ otherwise. 
Then $\mathcal{D}(S_A|\mathcal{G}(S_D))$ can be written as a function of $\mathbf{x}$ and $\mathbf{y}$:
\begin{align}
	\label{Eqn-attacker}
	F(\mathbf{x},\mathbf{y}) = \sum_{v_i \in V} (1-x_i)\cdot \min \{1, \sum_{v_j \in N^I(v_i)} y_j\}
\end{align}
where $N^I(v_i) = v_i \cup N(v_i)$. As a result, the defender's
problem \eqref{Eqn-bi-level} can be rewritten as
\begin{align}
	\label{Eqn-defender}
	&\min_\mathbf{x}\ \max_\mathbf{y} \ F(\mathbf{x},\mathbf{y})\\
	\text{s.t.}\quad & y_i \leq 1 - x_i,\ x_i, y_i \in \{0,1\},\ \forall i, \nonumber \\
	& \sum_{i = 1}^n x_i \leq k_D, \ \sum_{i = 1}^n y_i \leq k_A,\nonumber
\end{align}
where the first constraint ensures that the node blocked by
the defender cannot be selected as a seed by the attacker.

Next, we begin by developing a mixed-integer linear programming
formulation for the attacker's problem, and subsequently make use of
it to obtain both optimal and approximately optimal, but more scalable, solutions to the
defender's influence blocking problem.

\subsection{Computing Attacker's Best Response} We begin with the
attacker's problem. Fixing the defender's decision $\mathbf{x}$, the
attacker seeks to maximize the objective $F(\mathbf{x},\mathbf{y})$ in
\eqref{Eqn-attacker}.
We linearize each non-linear term $\min \{1, \sum_{v_j \in N^I(v_i)} y_j\}$ by replacing it with one auxiliary continuous variable $t_i \in [0,1]$ and one extra inequality $t_i \leq  \sum_{v_j\in N^I(v_i)} y_j$. Consequently, the attacker's problem can be formulated as a Mixed Integer Linear Program (with fixed $\mathbf{x}$), denoted as BR-MILP:
\begin{align}
\label{Eqn-MILP}
\max_{\mathbf{y},\mathbf{t}}\quad &\sum_{v_i \in V} (1-x_i)\cdot t_i \\
\text{s.t.}\quad &y_i \leq 1 - x_i,\ i = 1,2,\cdots,n \nonumber\\
&\sum_{v_i\in V} y_i \leq k_A,\ y_i \in \{0,1\} \nonumber \\
&t_i \leq  \sum_{v_j\in N^I(v_i)} y_j,\  0 \leq t_i \leq 1 \nonumber
\end{align}
The solution $\mathbf{y}^*$ to this MILP corresponds to the optimal strategy of the attacker given the defender's strategy $\mathbf{x}$.

\subsection{Optimal Influence Blocking: A Constraint Generation Approach}

\input{./sections/constraint}

\subsection{Approximating Optimal Influence}
The constraint generation approach enables us to effectively compute
optimal influence blocking.
However, it fails to scale to networks of even a moderate size.
We now propose a principled approximation approach that makes use of a
linear programming (LP) relaxation of the attacker's problem combined
with LP duality.

Specifically, by relaxing the integer constraint on each $y_i$, the
attacker's problem \eqref{Eqn-MILP} becomes a linear program (LP) with
variables $\mathbf{y}$ and $\mathbf{t}$.
Its dual is
\begin{equation}\label{dual3}
\begin{split}
\min_{\lambda_0,q,\alpha,\beta,\gamma}\quad  
&k_A\lambda_0 + \sum_{i=1}^n (1-x_i)q_i + \sum_{i=1}^{n} \beta_i+ \sum_{i=1}^{n} \gamma_i\\
\text{s.t.}\quad  & \lambda_0 + q_i + \beta_i -\sum_{v_j\in N^I(v_i)} \alpha_j \geq 0, \\
                  &\alpha_i+ \gamma_i \geq 1-x_i,\\
                  & \lambda_0, q_i,\alpha_i,\beta_i,\gamma_i \geq 0,  i = 1,2,\cdots,n
\end{split}
\end{equation}
where $\lambda_0, q, \alpha, \beta, \gamma$ are the dual variables.
By substituting the inner problem with \eqref{dual3}, the defender's bi-level program  can be reformulated as a minimization problem with the same objective as that in \eqref{dual3}, with the difference that $\mathbf{x}$ now are variables. 
Finally, we can linearize the non-linear terms $\sum_{i=1}^n
(1-x_i)q_i$ as follows.
We introduce new variables $w_i \geq 0$ and a large constant $M$, such that $w_i =(1-x_i)q_i$, $i = 1,2,\cdots,n$. 
We further introduce linear constraints for each $w_i$,$q_i$, and $x_i$:
\begin{align}
- M(1-x_i)  &\le w_i \leq M(1-x_i),\label{w1}\\
  q_i - Mx_i &\le w_i \leq q_i + Mx_i\label{w4}.
\end{align}

The full defender's problem can thus be formulated as a MILP, which we denoted by DEF-MILP:
\begin{equation}\label{final}
\begin{split}
\min_{x,\lambda_0,q,\mathbf{\alpha},\mathbf{\beta},\mathbf{\gamma},\mathbf{w}}\quad
& k_A\lambda_0 + \sum_{i=1}^n w_i + \sum_{i=1}^{n} \beta_i+ \sum_{i=1}^{n} \gamma_i\\
\text{s.t.}\quad &\sum_{i=1}^n x_i \leq k_D,\\
& \lambda_0 + q_i + \beta_i -\sum_{v_j\in N^I(v_i)} \alpha_j \geq 0,
\forall \ i\\
&\alpha_i+ \gamma_i \geq 1-x_i,\ \forall \ i\\
&\text{constraints} \quad (\ref{w1})-(\ref{w4})\\
&x_i \in \{1,0\}, w,\lambda_0, q,\alpha,\beta,\gamma \geq 0.
\end{split}
\end{equation}
The optimal strategy for the defender is then the solution
$\mathbf{x}^*$ to \eqref{final}.

\subsection{Scaling Up through a Pruning Heuristic}

\input{./sections/pruning}

%% file: sections/constraint.tex
We now propose a way to compute the exact solution to the bi-level
problem \eqref{Eqn-defender} by using a constraint generation method.
The defender's optimal problem can be alternatively expressed as the
following optimization problem:
\begin{align}
\min_{ \mathbf{x}, \mathbf{t}}\quad & \sum_{i = 1}^n t_i \\
\text{s.t.}\quad 
&\sum_{v_i\in V} x_i \leq k_D,\ x_i \in \{0,1\}  \\
&t_i = \min\{1-x_i, \sum_{j \in N^I(v_i)} \ y^*_j (1-x_j) \}, \forall i,\ \text{where} \label{min_}\\
&\ y^* = \text{BR-MILP}(x).
\end{align}
If we let 
$Y$ denote the complete set of the attacker's strategies, we can
further rewrite this
by a very large optimization problem in which we explicitly enumerate
all of the attacker's actions.
In this problem, the defender aims to find a strategy $\mathbf{x}$ such that
the tight upper bound of the attacker's utility is minimized. For each
$\mathbf{y} \in Y$, we can introduce the corresponding variables $t_{i,y}$
showing whether node $i$ is influenced given the attacker's strategy
$\mathbf{y}$.
Constraint \eqref{min_} given each $\mathbf{y}$ can be linearized to
\eqref{m_head} - \eqref{m_tail} by introducing binary variables
$b_{i,y}$ which indicates whether $1-x_i < \sum_{j \in N^I(v_i)} \ y_j
(1-x_j)$.
Introducing a sufficiently large constant $M$ allows us to further
linearize all of the non-linear terms, yielding the following:
%
\begin{align}
\label{DEF-MASTER}
\min_{ \mathbf{x},\mathbf{t^d}}\quad &U_A \\
\text{s.t.}\quad 
&\sum_{v_i\in V} x_i \leq k_D,\ x_i \in \{0,1\}  \\
&U_A \geq \sum_{i = 1}^n t^d_{i, y}, \forall y \in Y \label{m_head}\\ 
& 1 - x_i - M(1 - b_{i,y})\leq t_{i,y} \leq 1 - x_i, \forall y \in Y,\ \forall  i\label{m}\\
& \sum_{j \in N^I(v_i)} y_j (1-x_j)  - Mb_{i,y}\leq t_{i,y} \leq \sum_{j \in N^I(v_i)} \ y_j (1-x_j), \forall y \in Y,\ \forall  i \label{m_tail}
\end{align}

However, the MILP above is clearly intractable since the set $Y$ is
combinatorial.
To tackle the computational issue, we develop a constraint generation
algorithm.
The key to this algorithm is to replace $Y$ with a small subset of
attacker strategies $\hat Y \subset Y$, along with all of the
associated constraints, so that the modified MILP
above becomes DEF-MASTER($\hat{Y}$), in which we can specify an
arbitrary subset of attacks $\hat{Y}$.
Now we can start by an arbitrary small set of attacks, and
interleave two steps: solve DEF-MASTER($\hat{Y}$) using the set of
attacks $\hat{Y}$ generated so far to obtain a provisional solution
$\mathbf{x}$ for the defender, and identify a new attack $\mathbf{y}$ that
is a best response to $\mathbf{x}$.
We can stop this as soon as the best response of the attacker no
longer improves their utility compared to the solution obtained by
DEF-MASTER($\hat{Y}$).
Algorithm~\ref{cg} fully formalizes the proposed constraint generation
procedure, where is the set of optimal  BR-MILP is just the mixed-integer linear programming
approach for identifying the best response of the attacker presented
in formulation~\eqref{Eqn-MILP}.
Note that we can utilize the returned influence value $t^a_y$ of
BR-MILP to prune irrelevant constraints of DEF-MASTER. Specifically,
we only generate constraints \eqref{m} - \eqref{m_tail} for each
influenced node ($t^a_{i,y} = 1$). For the node with $t^a_{i,y} = 0$,
we add the constraint $t^d_{i,y} = 0$, because given an attacker's
stratgy $\mathbf{y}$, the uninfluenced node will not be influenced no matter
what $\mathbf{x}$ is. Consequently, we denote the refined master problem by
DEF-MASTER($\hat Y$, $\hat T^a$) in Algorithm~\ref{cg}.


\begin{algorithm}[h]
	\caption{Constraint Generation (CG)}
	\label{cg}
	\begin{algorithmic}[1]      
		\State $\hat Y = \emptyset$, $\hat T^a =  \emptyset$, 
		\State $U_A^{UB} = \infty$, $U_A^{UB} = 0$
		\State $x^*, x_{def} = \vec{0}$
		\While {$U_A^{UB} - U_A^{LB} > gap$}
		\State $(t^a_y, y, U_A) \leftarrow$ BR-MILP($x_{def}$)
		\State $\hat Y = \hat Y \cup \{y\}$, $\hat T^a = \hat T^a \cup \{t^a_{y}\}$
		\If {$U_A < U_A^{UB}$}
		\State Update the upper bound $U_A^{UB} = U_A$
		\State Update the incumbent solution $x^* \leftarrow x_{def}$
		\EndIf
		\State $(x_{def}, U_A^{LB}) \leftarrow$ DEF-MASTER($\hat Y$, $\hat T^a$)
		\EndWhile
		\State \textbf{return} $x^*$
	\end{algorithmic}
\end{algorithm}


%% file: sections/pruning.tex

\begin{algorithm}[h]
	\caption{Heuristic Pruning Algorithm}
	\label{alg-pruning}
	\begin{algorithmic}[1]
		\Procedure{Pruned-MILP}{$k_A$, $k_D$, $l_D$, $G = (V,E)$}       
		\State SortedList = SortingAlg($V$) 
		\State $\mathbb{X}_{pruned} \leftarrow  \{0,1\}^n$
		\For{$i$ in SortedList[$l_d$, $n$]} 
		\State $\forall \mathbf{x}\in \mathbb{X}_{pruned}$, fix $\mathbf{x}[i] = 0$
		\EndFor
		\Comment{Limit the strategy space to top $l_d$ nodes}
		\State $x_{def}$ $\leftarrow$ DEF-MILP($\mathbb{X}_{pruned}$, $k_A$, $k_D$, $G$)
		\State $lastNum \leftarrow k_d$ - \Call{calBlockedNum}{$x_{def}$} 
		\For{$i$ in SortedList} 
		\If{$lastNum \leq 0$}
		\State \textbf{break}
		\EndIf
		\If{$x_{def}[i] == 0$}
		\State $x_{def}[i] =  1$
		\State $lastNum = lastNum - 1$
		\EndIf
		\EndFor
		\State \textbf{return} $x_{def}$
		\EndProcedure
		\Procedure{calBlockedNum}{$x_{def}$}  
		\State $num = 0$
		\For{$i$ in $x_{def}$} 
		\State $num += i$
		\EndFor
		\State \textbf{return} $num$
		\EndProcedure
	\end{algorithmic}
\end{algorithm}
Even finding the approximately optimal strategy for the defender above involves solving a MILP \eqref{final}, of which the number of constraints grows linearly with the number of nodes. 
This is a computational bottleneck, especially when the network is large. 
We propose a heuristic approach to deal with very large networks. 
The basic idea is to limit the strategy space of the defender.

We write the DEF-MILP \eqref{final} as a function DEF-MILP($\mathbb{X}$, $k_A$, $k_D$, $\mathcal{G}$), where $\mathbb{X}$ denotes the strategy space of the defender. 
Our algorithm relies on pruning some \textit{less important} nodes, which significantly reduce the strategy space $\mathbb{X}$. 
Note that the importance of the nodes can be measured by different metrics, such as the node degree.
Our Heuristic Pruning Algorithm is presented in Alg.~\ref{alg-pruning}. 
The idea is to first sort the nodes according to some importance
metric in descending order, and then restrict the defender's strategy space in the top-$l_D$ nodes; that is, setting $x_i = 0$ for the rest. 
Finally, we solve the MILP with restricted strategy space. 
The parameter $l_D$ controls the trade-off between the time complexity of solving the MILP and the quality of the solution.

%% file: sections/extensions.tex
\section{Extensions}

\paragraph{Weighted Influence Maximization}

A natural extension of \emph{influence maximization} allows each node
$v_i\in V$ to be associated with non-negative weight $\mu_i$ capturing its importance in the final outcome \cite{Kempe:2003:MSI:956750.956769}. 
Here we denote this problem as \emph{weighted influence maximization} (WIM).
They defined the weighted influence function $\sigma_{\mu}(S)$ as the expected value outcomes $B$ of the quantity $\sum_{v_i\in B} \mu_{i}$, where $B$ denotes the random set activated by the process with initial seed set $S$.

To incorporate weighted influence maximization, we generalize our model by associating a weight to each node in the objective function \eqref{Eqn-attacker}, i.e.,  $F(\mathbf{x},\mathbf{y}) = \sum_{v_i \in V} \mu_i(1-x_i)\cdot \min\{1, \sum_{v_j\in N^I(v_i)} y_j\}$. 

The inner problem of the attacker's best response can be formulated by modifying the objective in (\ref{Eqn-MILP}) to $\sum_{v_i \in V} \mu_i (1-x_i)\cdot t_i$.

Applying the same procedure of calculating the defense strategy of the
non-weighted version, we can formulate the defender's optimization
problem. The procedure is briefly described as follows. First, we can
directly generalize the MILP formulation of the attacker's best response.
Next, we relax the integer constraint on each $y_i$ and take the dual
of the resulting LP.
The bi-level problem can then be reformulated as a non-linear minimization problem by replacing the inner problem with the relaxed dual. 
Finally, we introduce the large number $M$ to linearize the non-linear term, we can get the final formulation, denoted as DEF-WMILP, shown as follows.

\begin{subequations}
  \label{weightedMILP}
\begin{align}
\min_{x,\lambda_0,q,\mathbf{\alpha},\mathbf{\beta},\mathbf{\gamma}}\ \ 
& k_A\lambda_0 + \sum_{i=1}^n w_i + \sum_{i=1}^{n} \beta_i+ \sum_{i=1}^{n} \gamma_i\\
s.t.\ \ \ & w,\lambda_0, q,\alpha,\beta,\gamma \geq 0\\
&\sum_{v_i\in V} x_i \leq k_D\\
& \lambda_0 + q_i + \beta_i -\sum_{v_j\in N^I(v_i)} \alpha_j \geq 0, \forall i\\
&\alpha_i+ \gamma_i \geq \mu_i(1-x_i), \forall i\\
&\text{constraints} \quad (\ref{w1})-(\ref{w4})\\
&x_i \in \{1,0\}, \forall i.
\end{align}
\end{subequations}
For the heuristic pruning algorithm \Call{Pruned-MILP}{}, we can substitute DEF-MILP with DEF-WMILP. 

\paragraph{Blocking both Edges and Nodes}

The model can be further generalized by considering blocking both
edges and nodes with different costs. Suppose that the cost of
blocking an edge is $c_e$ and the cost of blocking a node is $c_n$,
and the defender chooses to block a subset of both edges and nodes
given a total budget $B_D$.
Let $z_{ij} = \{0,1\},\ \forall (i,j)\in E$ denote the defender's edge strategy, where $z_{ij} = 1$ means that the defender chooses to block edge $(i,j)$ and $z_{ij} = 0$ otherwise. Then the defender's budget constraint becomes:
\begin{align}
	\label{budget_cons}
	\sum_{v_i \in V} x_ic_n + \sum_{(i,j) \in E}z_{ij}c_e \leq B_D
\end{align}

Once blocking a node, it is not necessary to block the edges linked to the node. To demostrate this node-edge relationship, we introduce an integer variable $k_{ij}\in \{0,1\},\ \forall (i,j)\in E$ and the following linear constraints.

\begin{align}
& z_{ij} - 0.5 \leq Mk_{ij} \label{edge_node1}\\
& x_i - 0.5 \leq M(1-k_{ij}) \\
& x_j - 0.5 \leq M(1-k_{ij}) \label{edge_node3}
\end{align}

Given the defender's strategy $\mathbf{z}$ and $\mathbf{x}$, the attacker's best response can be modified to 
\begin{align}
\label{EV-MILP}
\max_{ \mathbf{y},\mathbf{t}}\quad &\sum_{v_i \in V}  (1-x_i)\cdot t_i \\
\text{s.t.}\quad 
&\sum_{v_i\in V} y_i \leq k_A,\ y_i \in \{0,1\}  \\
&y_i \leq 1 - x_i, \forall i\\ 
&t_i \leq  y_i + \sum_{v_j\in N(v_i)} y_j(1-z_{ji}),\  0 \leq t_i \leq 1, \forall i
\end{align}
Finally, taking the dual of the relaxed attacker's problem, the defender's problem can be formulated as a non-linear mixed integer program:
\begin{equation}\label{EV-MIP}
\begin{split}
\min_{x,Z,\lambda_0,q,\mathbf{\alpha},\mathbf{\beta},\mathbf{\gamma}}\ \ 
& k_A\lambda_0 + \sum_{i=1}^n (1-x_i)q_i + \sum_{i=1}^{n} \beta_i+ \sum_{i=1}^{n} \gamma_i\\
s.t.\ \ \ & w,\lambda_0, q,\alpha,\beta,\gamma \geq 0\\
&\sum_{v_i \in V} x_ic_n + \sum_{(i,j) \in E}z_{ij}c_e \leq B_D\\
& \lambda_0 + q_i + \beta_i -\alpha_i -\sum_{v_j\in N(v_i)} \alpha_j(1-z_{ij}) \geq 0, \forall i\\
&\alpha_i+ \gamma_i \geq \mu_i(1-x_i), \forall i\\
&\text{constraints} \quad (\ref{edge_node1})-(\ref{edge_node3})
\end{split}
\end{equation}

We can linearize the non-linear terms by replacing $(1-x_i)q_i$ by
introducing a new variable $w_i$ and replacing $\alpha_j(1-z_{ij})$
with $b_{ij}$. Then the optimal defense strategy
$(\mathbf{x}^*,\mathbf{z^*})$ can be obtained by solving the
large-scale MILP (\ref{EV-MILP}). 
\begin{subequations}
  \label{EV-MILP2}
\begin{align}
\min_{x,Z,\lambda_0,q,\mathbf{\alpha},\mathbf{\beta},\mathbf{\gamma},\mathbf{b},\mathbf{k},\mathbf{w}}\ \ 
& k_A\lambda_0 + \sum_{i=1}^n w_i + \sum_{i=1}^{n} \beta_i+ \sum_{i=1}^{n} \gamma_i\\
s.t.\ \ \ & w,k,b,\lambda_0, q,\alpha,\beta,\gamma \geq 0\\
&\sum_{v_i \in V} x_ic_n + \sum_{(i,j) \in E}z_{ij}c_e \leq B_D\\
& \lambda_0 + q_i + \beta_i -\alpha_i -\sum_{v_j\in N(v_i)} b_{ij} \geq 0, \forall i\\
&\alpha_i+ \gamma_i \geq \mu_i(1-x_i),\forall i\\
& M(1-z_{ij}) \geq b_{ij} \geq - M(1-z_{ij}), \forall (i,j) \in E\\
&\alpha_j + Mz_{ij} \geq b_{ij} \geq \alpha_j - Mz_{ij}, \forall (i,j) \in E\\
  &\text{constraints} \quad (\ref{w1})-(\ref{w4}), \quad (\ref{edge_node1})-(\ref{edge_node3})\nonumber\\
     &z_{ij}, k_{ij} \in \{0,1\}, \forall (i,j)\in E; x_i \in \{0,1\}, \forall i. \nonumber
\end{align}
\end{subequations}

%% file: sections/exp.tex
\section{Experiments}
In this section, we test our defense approaches against several
attacks and also compare them with several defense baselines from
previous works.
All runtime experiments were performed on a 2.6 GHz 8-core Intel Core
i7 machine with 16 GB RAM.
The MILP instances were solved using CPLEX version 12.10.

\begin{figure*}[ht] 
\centering
\includegraphics[scale=0.35]{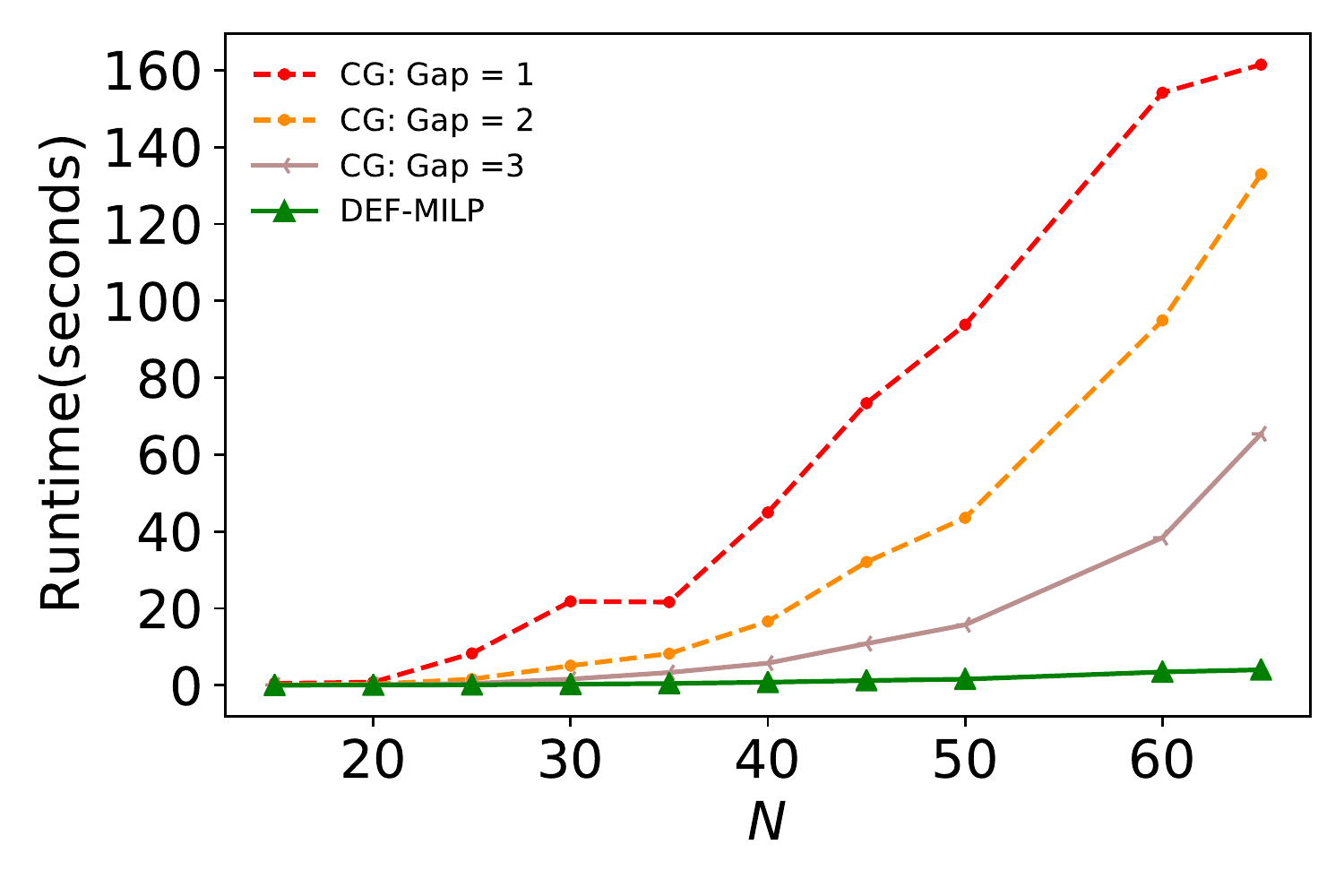}
\includegraphics[scale=0.35]{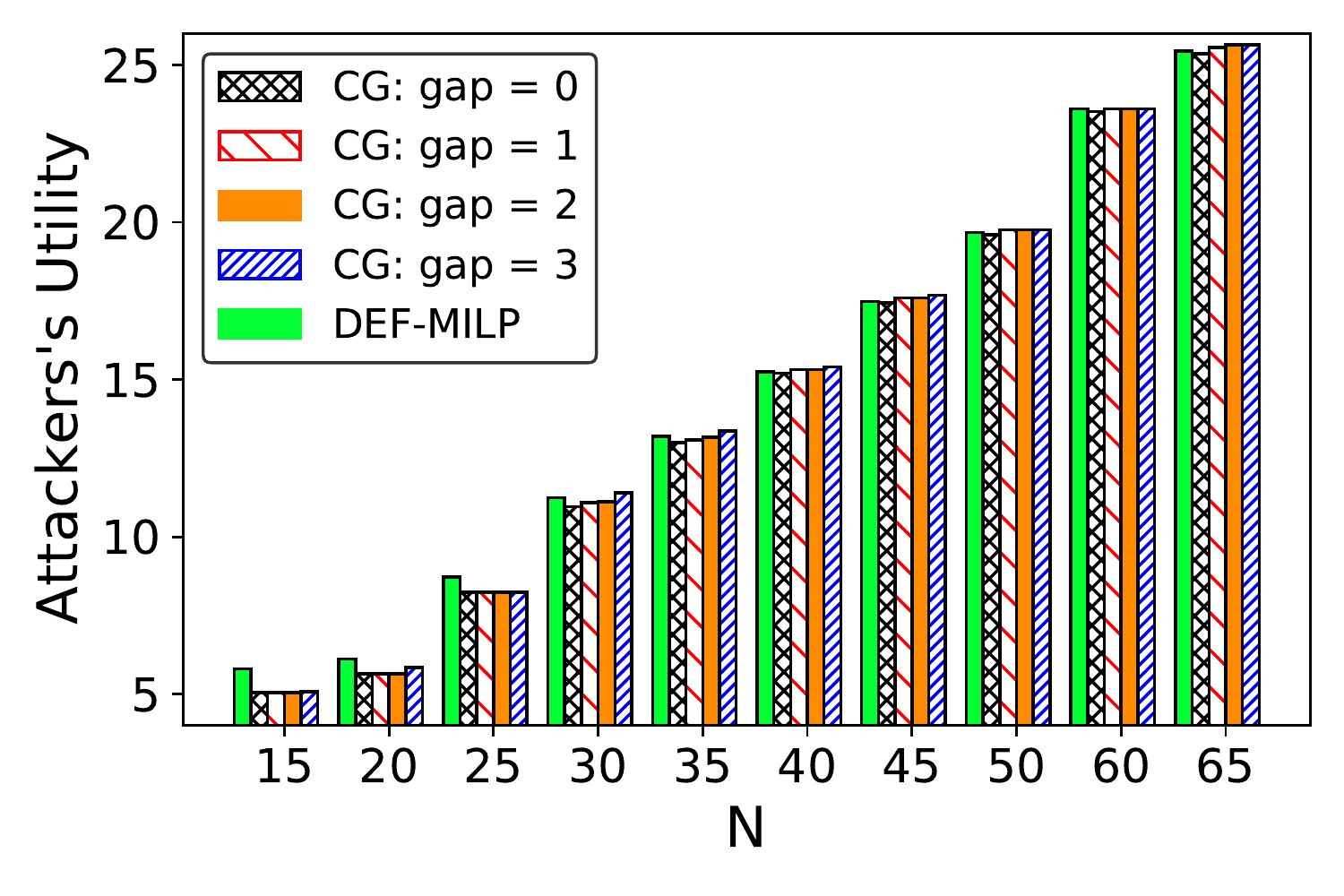}
\caption{Comparison between Constraint generation and DEF-MILP in terms of runtime(left) and the attacker's utility (right).} 
\label{fig:cg}
\end{figure*}


\begin{figure*}[ht] 
\centering
\subfigure[ER (k-MaxVD)]{
\includegraphics[scale=0.18]{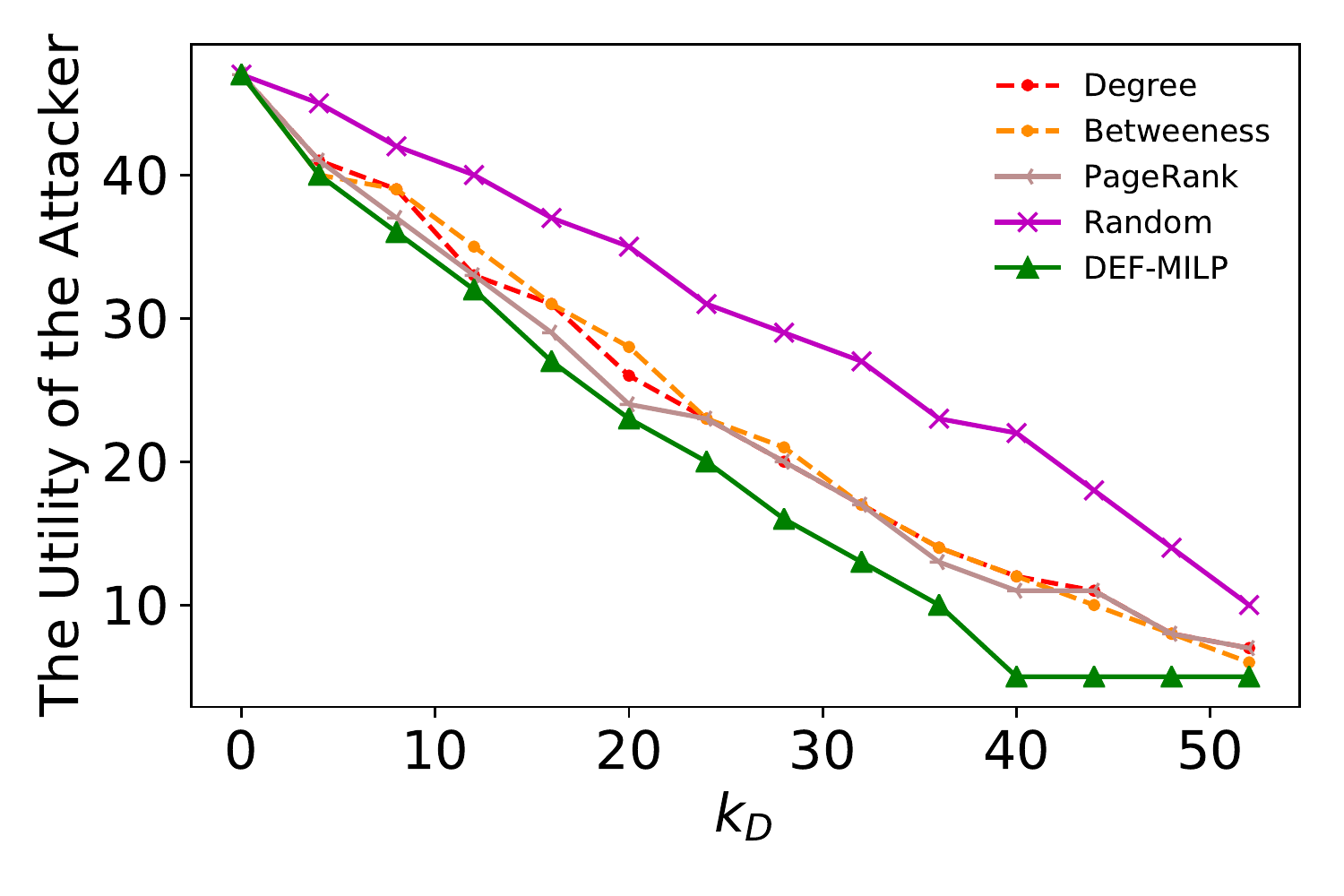}
}
\subfigure[ER (IC)]{
\includegraphics[scale=0.18]{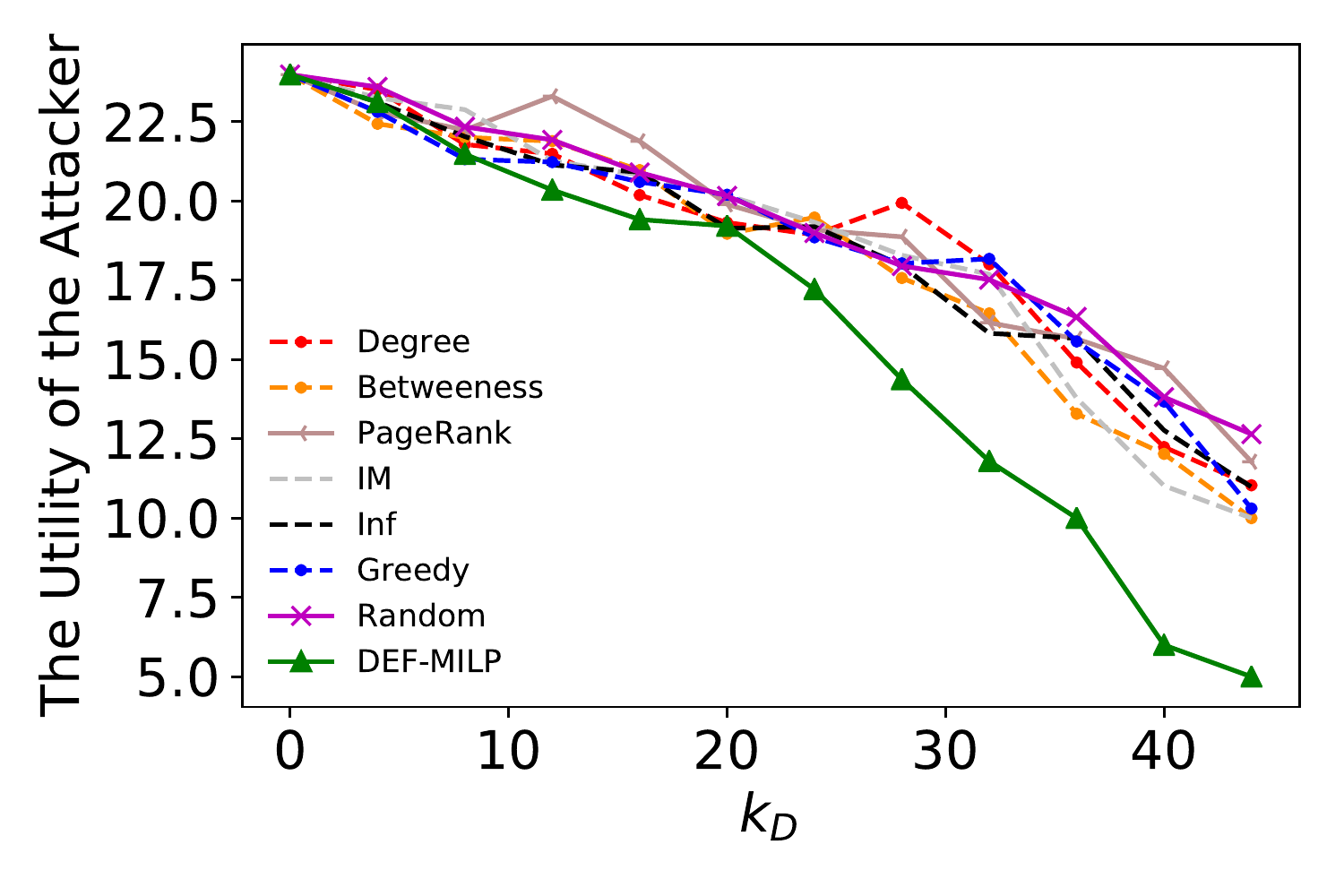}
}%
\subfigure[ER (LT)]{
\includegraphics[scale=0.18]{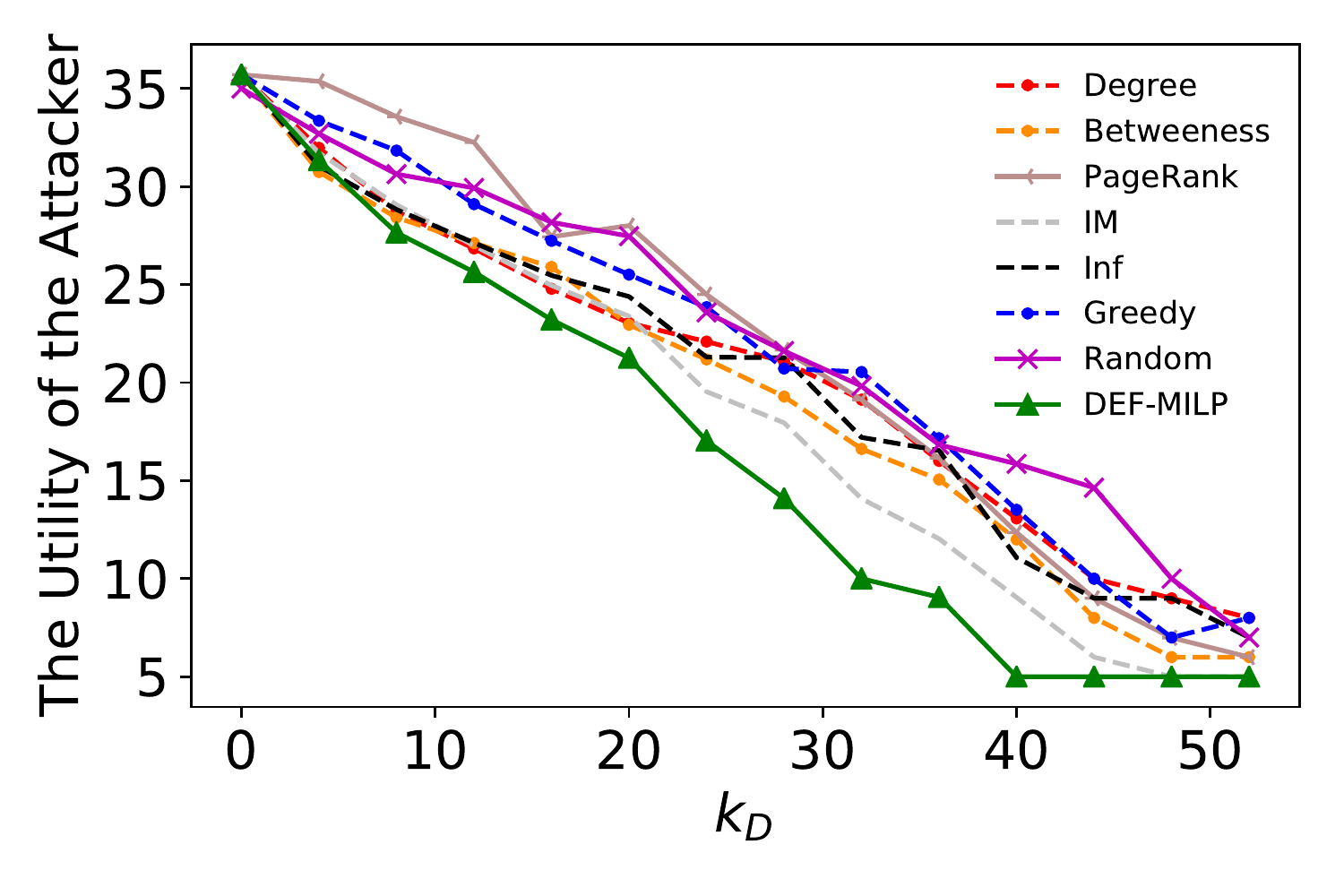}
}%
\subfigure[ER (WIM)]{
\includegraphics[scale=0.18]{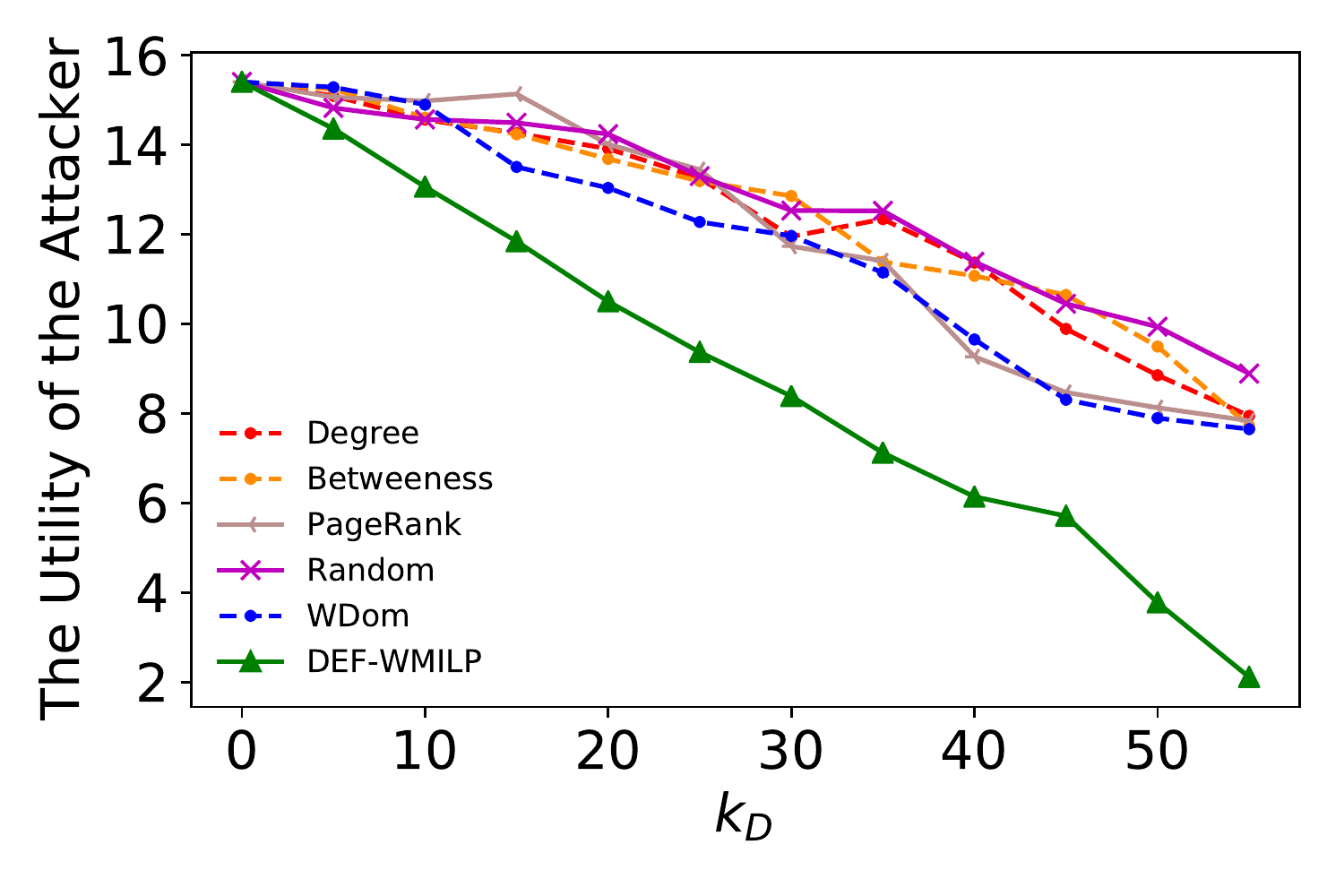}
}%

\subfigure[WS (k-MaxVD)]{
\includegraphics[scale=0.18]{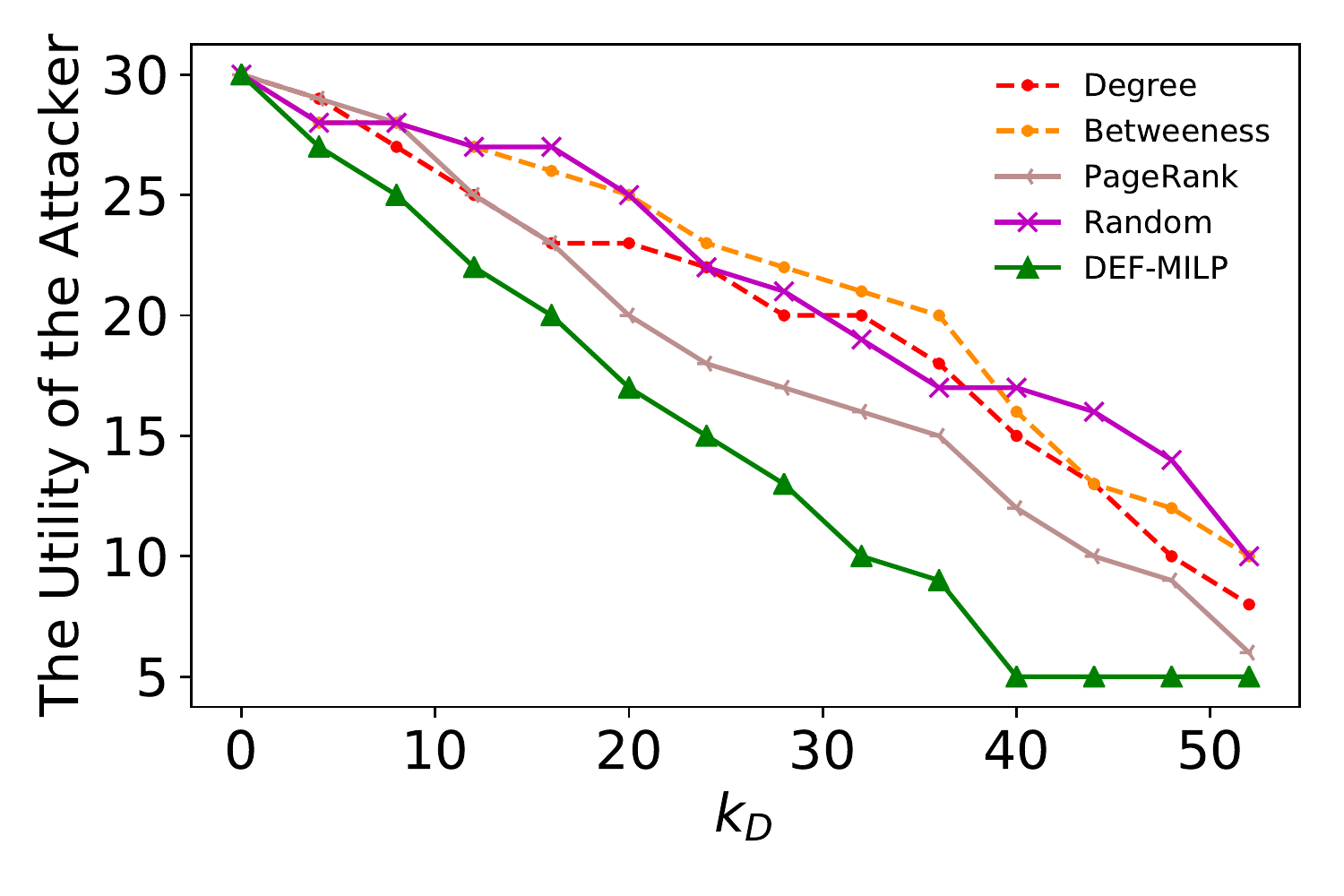}
}%
\subfigure[WS (IC)]{
\includegraphics[scale=0.18]{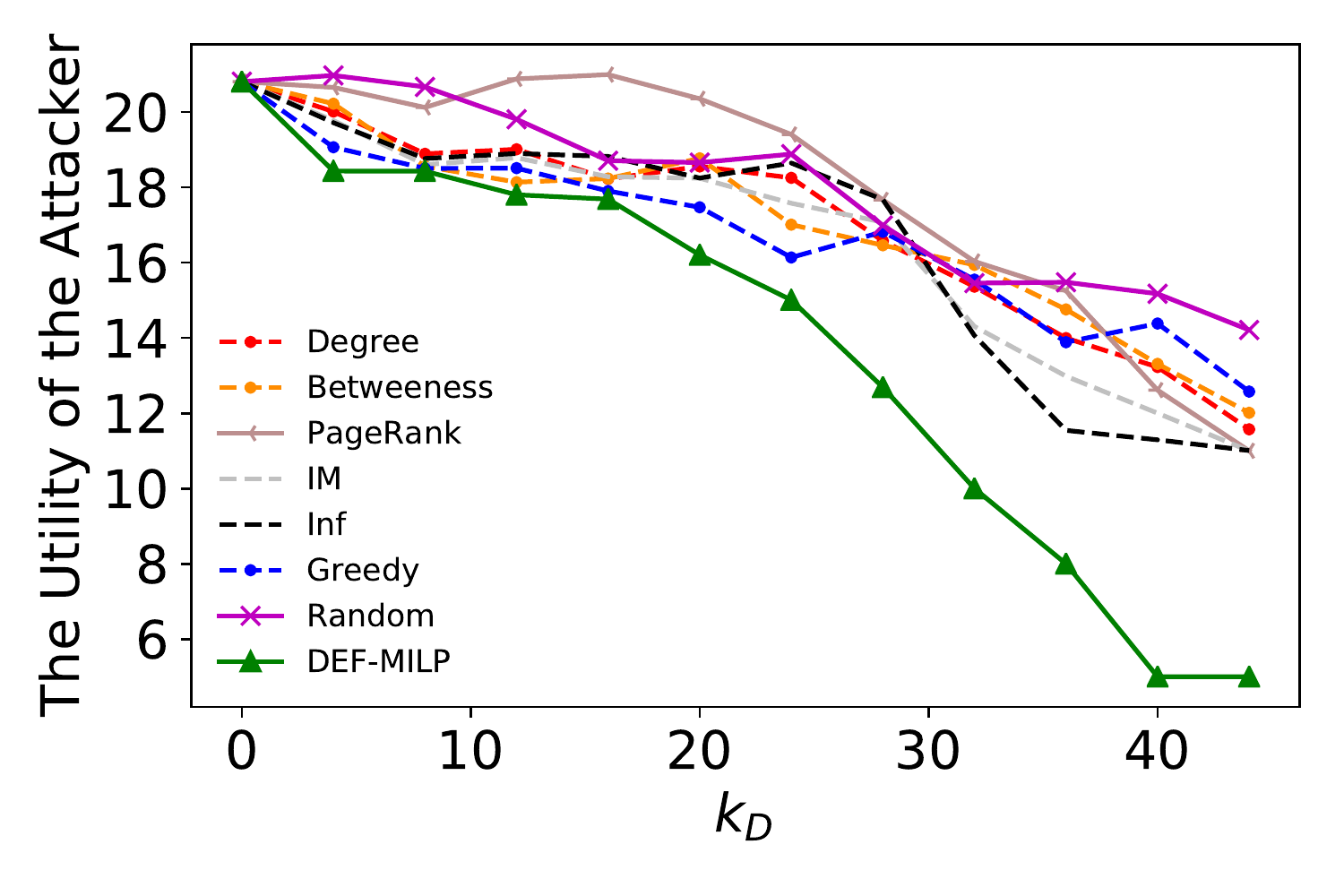}
}%
\subfigure[WS (LT)]{
\includegraphics[scale=0.18]{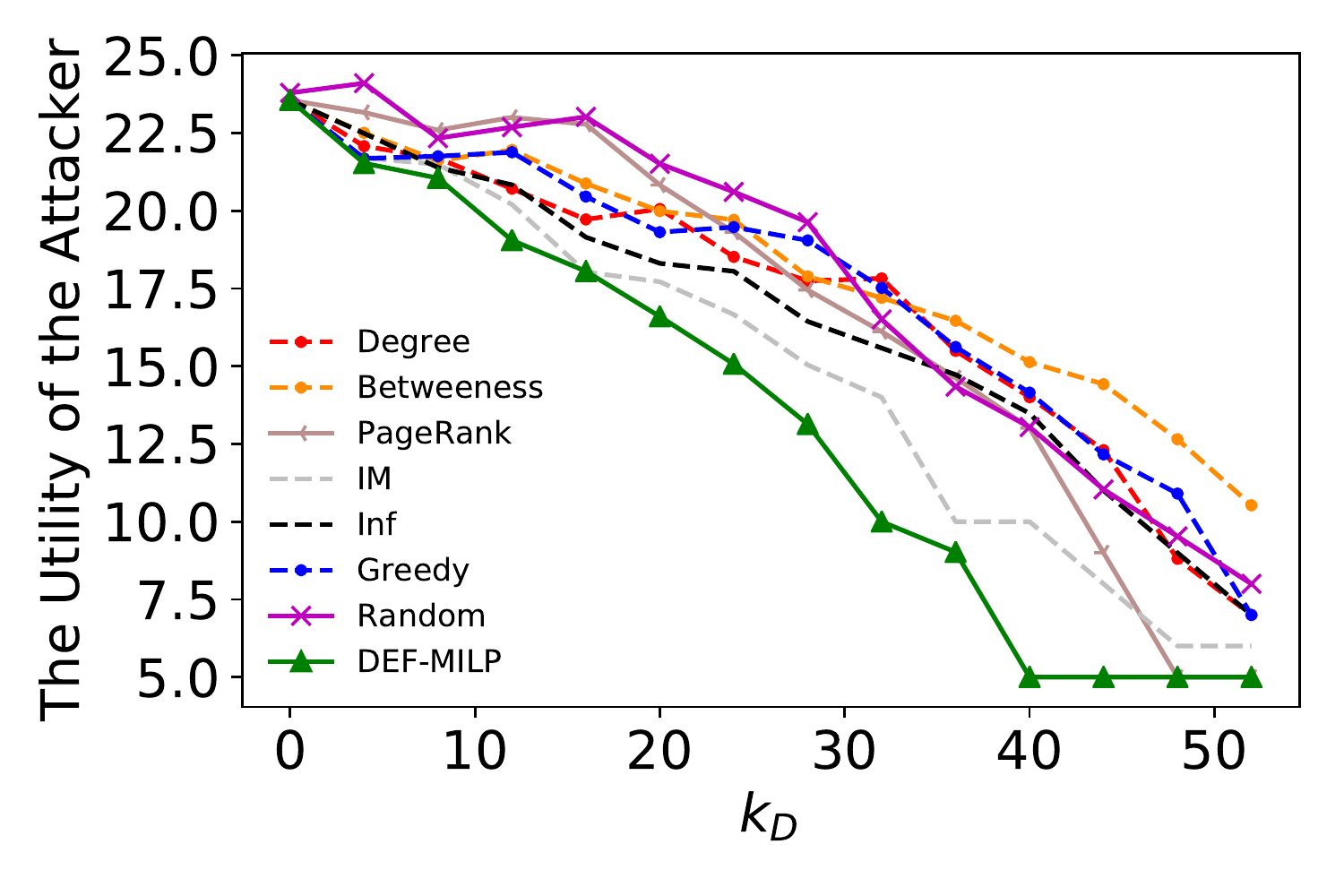}
}%
\subfigure[WS (WIM)]{
\includegraphics[scale=0.18]{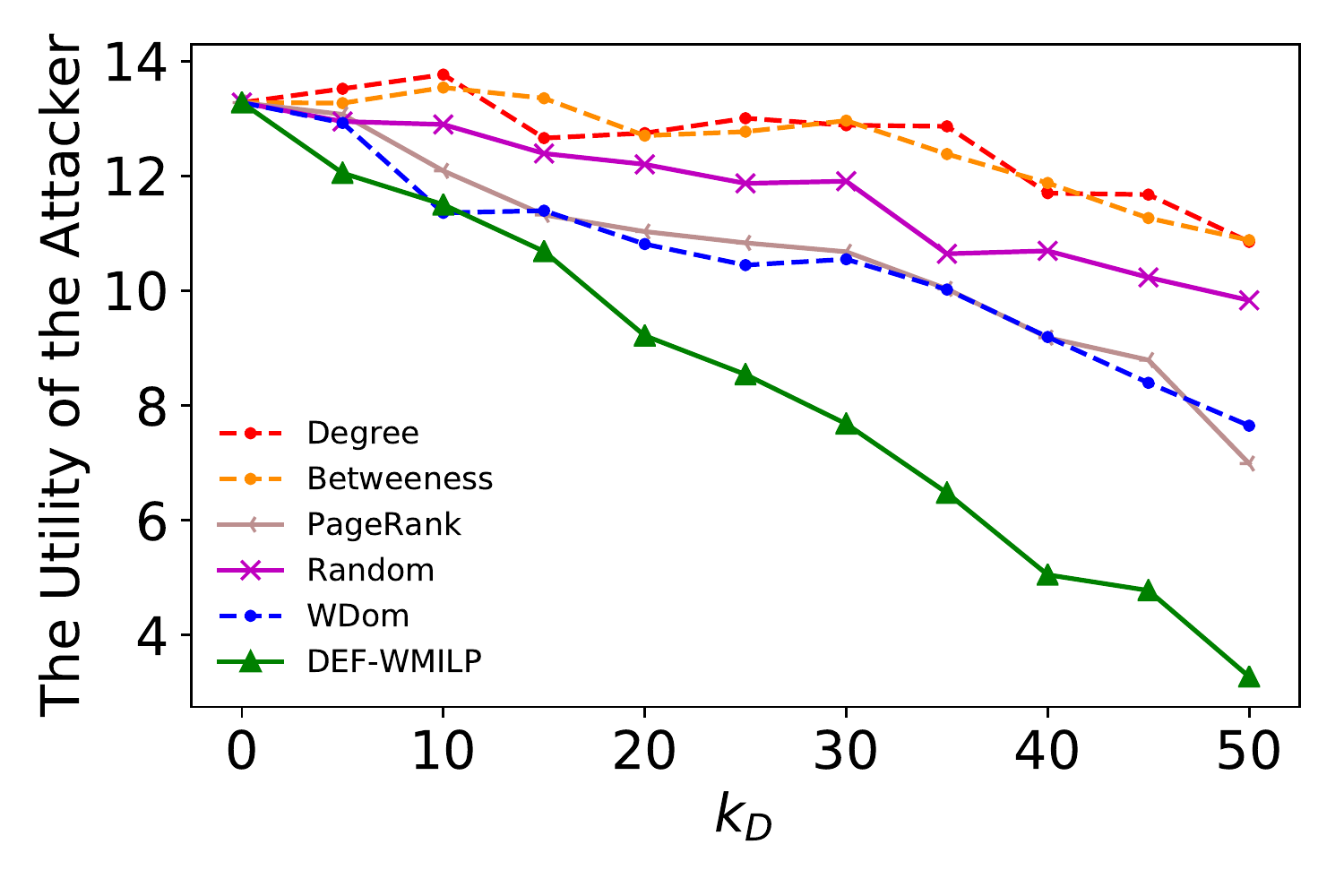}
}%

\subfigure[BA (k-MaxVD)]{
\includegraphics[scale=0.18]{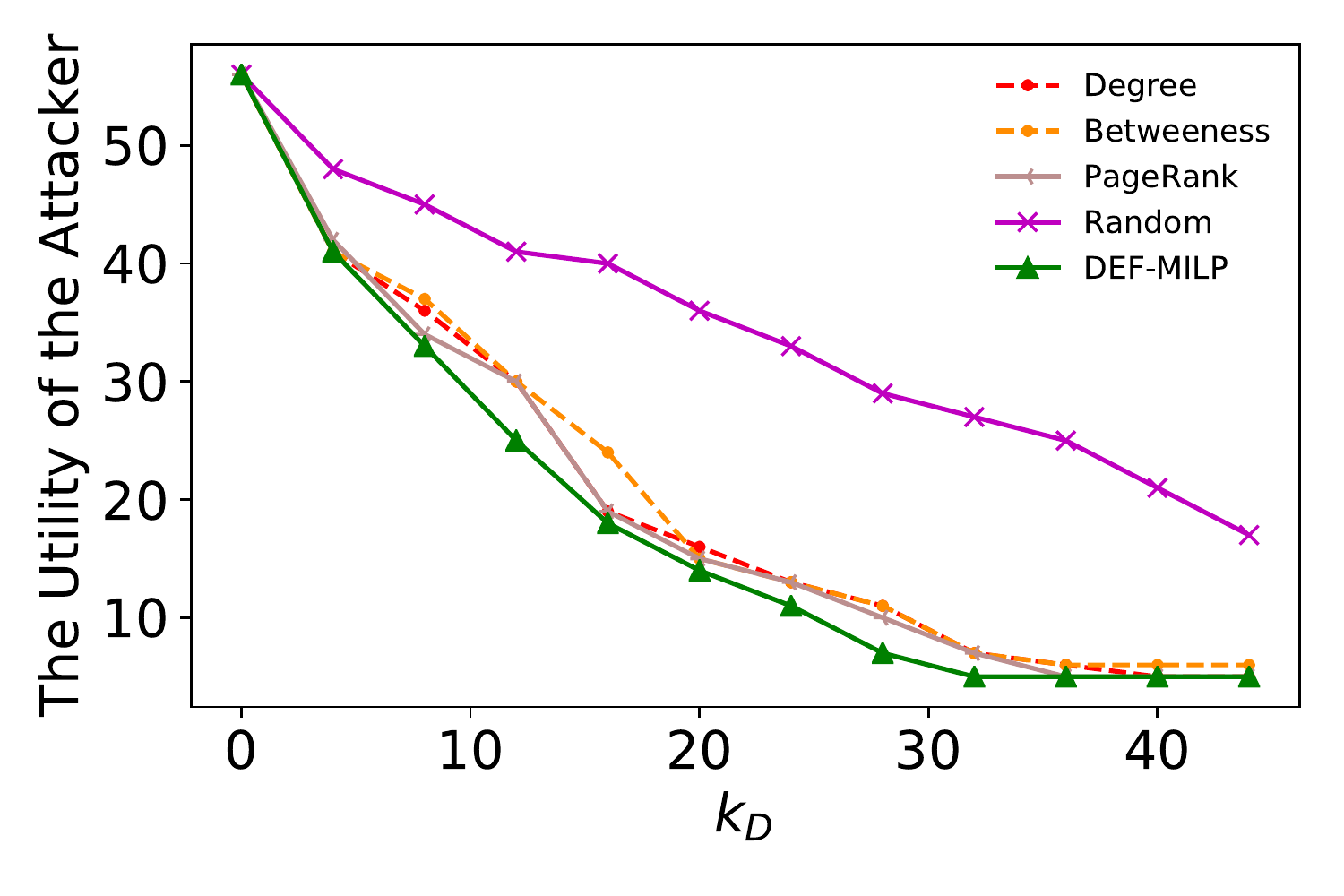}
}%
\subfigure[BA (IC)]{
\includegraphics[scale=0.18]{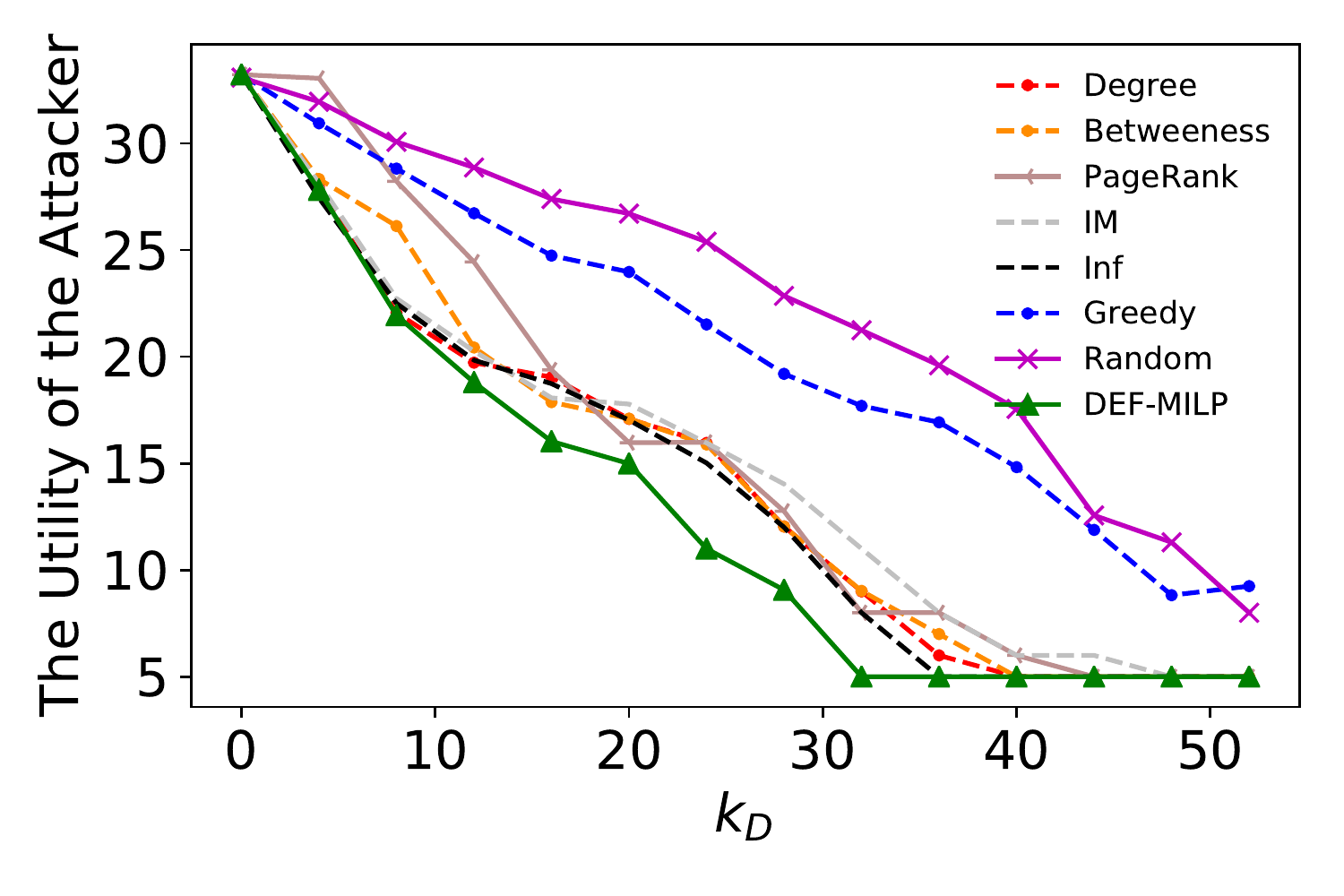}
}%
\subfigure[BA (LT)]{
\includegraphics[scale=0.18]{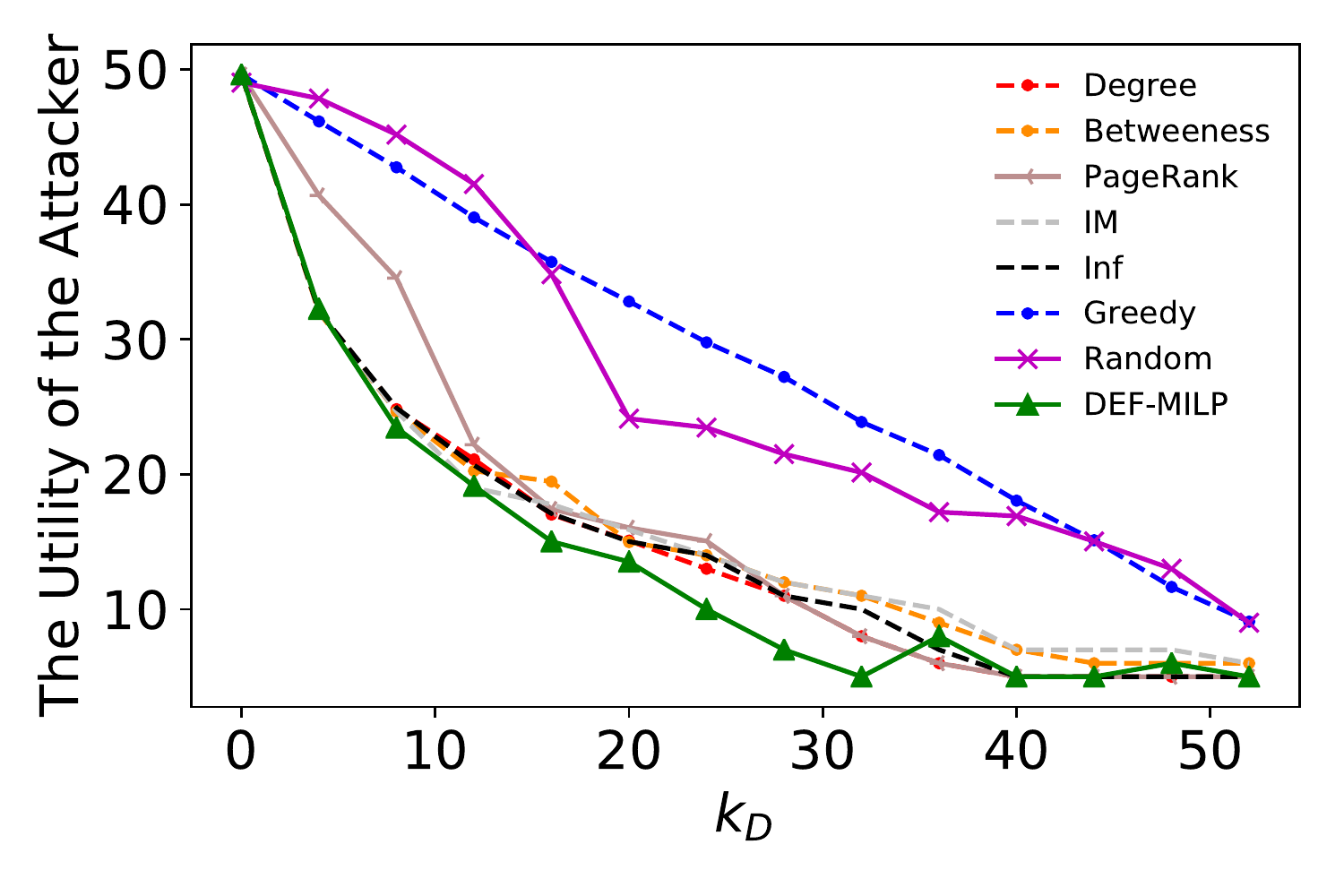}
}%
\subfigure[BA (WIM)]{
\includegraphics[scale=0.18]{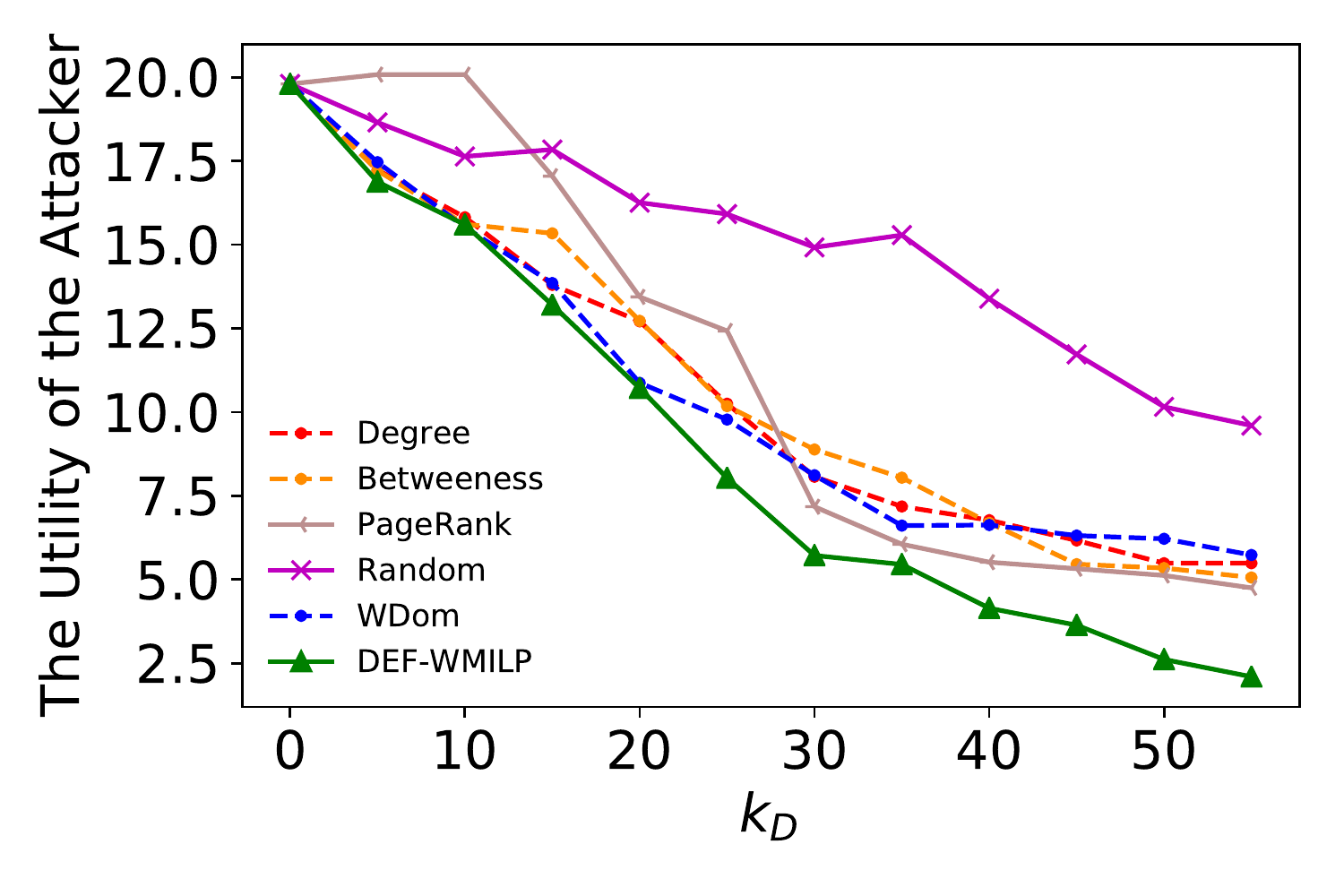}
}%

\caption{The performance of DEF-MILP on synthetic networks against several attacks.} 
\label{random}
\end{figure*}

\paragraph{Data sets} We conduct experiments on both synthetic graphs and real-world networks.
\begin{itemize}
    \item \textbf{Synthetic graphs}: We generate synthetic graphs from three graph models: Erdos-Renyi (ER) model~\cite{erdos59a}, Watts-Strogatz (WS) model~\cite{watts1998collective} generates networks with small-world properties., and Barabasi-Albert (BA) model~\cite{watts1998collective}. Specifically, each edge in the ER model is generated with probability $p = 0.1$. In the WS model, each node is initially connected to $5$ nodes in a ring topology and we set the rewiring probability as $0.15$. 
    In the BA model, at each time we add a new node with $m=3$ links that attach to existing nodes.

    \item \textbf{Real-world networks}: 
    We consider four real-world networks.  
    The Email-Eu-Core network \cite{Yin:2017:LHG:3097983.3098069,Leskovec:2007:GED:1217299.1217301} is generated using email data from a large European research institution, which has 1,005 nodes and 25,571 edges. 
    The Hamsterster friendships network \cite{konect:2016:petster-friendships-hamster,konect} is an undirected friendship network of the website \emph{hamsterster.com} with 1,858 nodes and 12,534 edges. 
    We also tested on the sub-networks of 
    a Facebook friendship network \cite{McAuley:2012:LDS:2999134.2999195} and 
    the Enron email network \cite{DBLP:journals/corr/abs-0810-1355,klimt04introducing}, where the sub-networks are sampled by the Forest Fire sampling method \cite{Leskovec:2006:SLG:1150402.1150479}.
\end{itemize}

\paragraph{Methodology}  Given a graph $\mathcal{G}$, we employ a defense strategy to block $k_D$ nodes, resulting in a modified graph $\mathcal{G}^M$. The attacker then uses an attack strategy to select $k_A$ seeds to spread the influence. We then measure the utility of the attacker under various combinations of defense and attack strategies. Specifically, we test our proposed defense strategies (CG, DEF-MILP, DEF-WMILP, and the corresponding pruned algorithms) against three attacks (\texttt{k-MaxVD}, \texttt{IM}, \texttt{WIM}).
We also compare our defense strategies with several defense baselines. These attack and defense strategies are detailed as follows.

\paragraph{Attacks}
We consider three types of attacks: \texttt{k-MaxVD}, \texttt{IM}, and \texttt{WIM}. In the  \texttt{k-MaxVD} attack, the attacker solves BR-MILP \eqref{Eqn-MILP} to find the seeds.
In the \texttt{IM} attack, the attacker employs an efficient variation, termed \emph{CELF-greedy}  \cite{Leskovec:2007:COD:1281192.1281239}, of the classical greedy algorithm \cite{Kempe:2003:MSI:956750.956769} to solve the influence maximization problem. Specifically, \emph{CELF-greedy} utilizes the submodularity of the spread function and conduct an early termination heuristic, which achieves up to 700 times efficiency improvement while still providing a $(1-1/e)$ approximation guarantee. The \texttt{WIM} attack is a variation of the \texttt{IM} attack adapted to the weighted setting.

\begin{figure} 
	\centering
	\subfigure[Hamsterster($k_A = 30$)]{
		\includegraphics[scale=0.3]{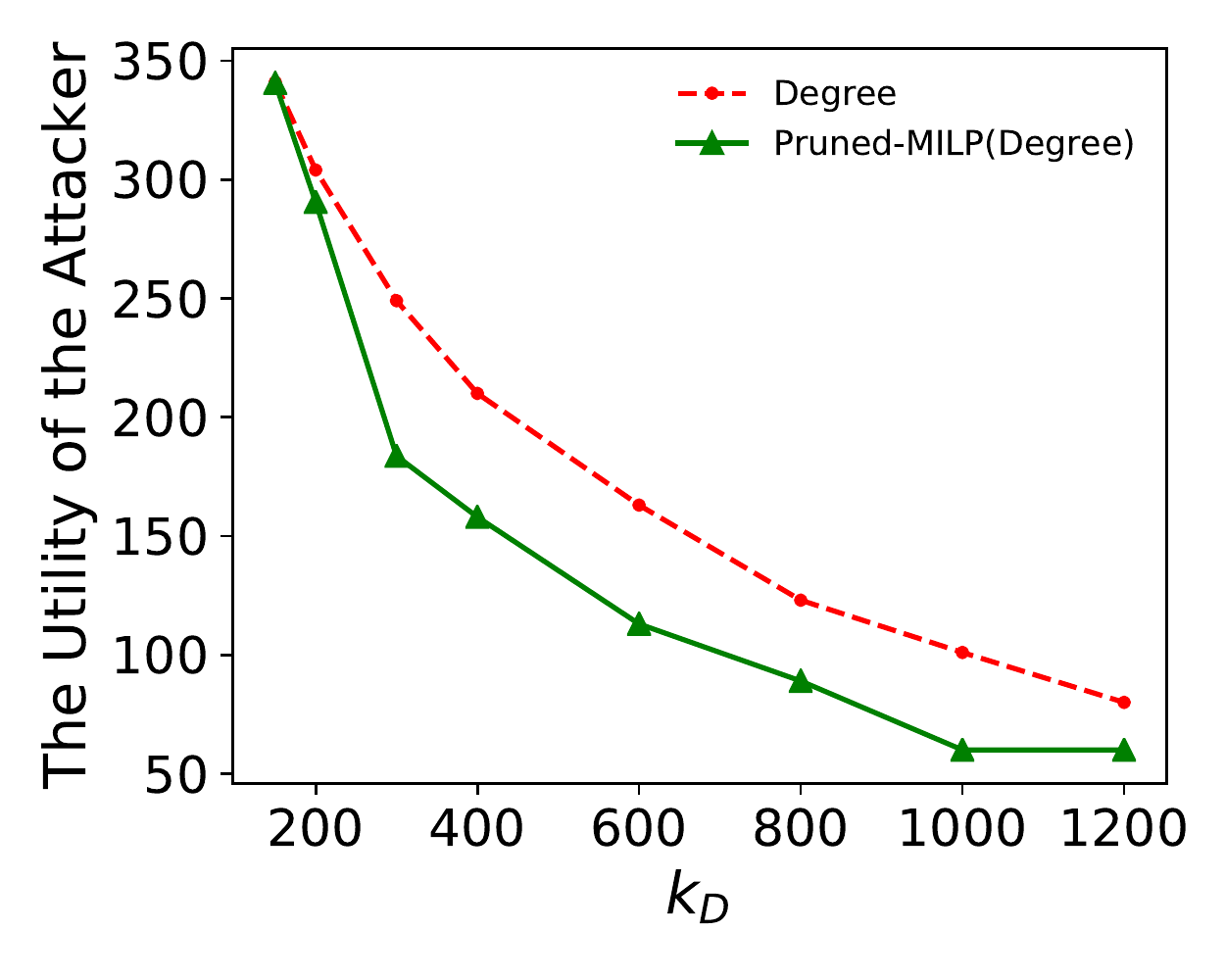}
	}%
	\subfigure[Email-Eu-core($k_A = 20$)]{
		\includegraphics[scale=0.3]{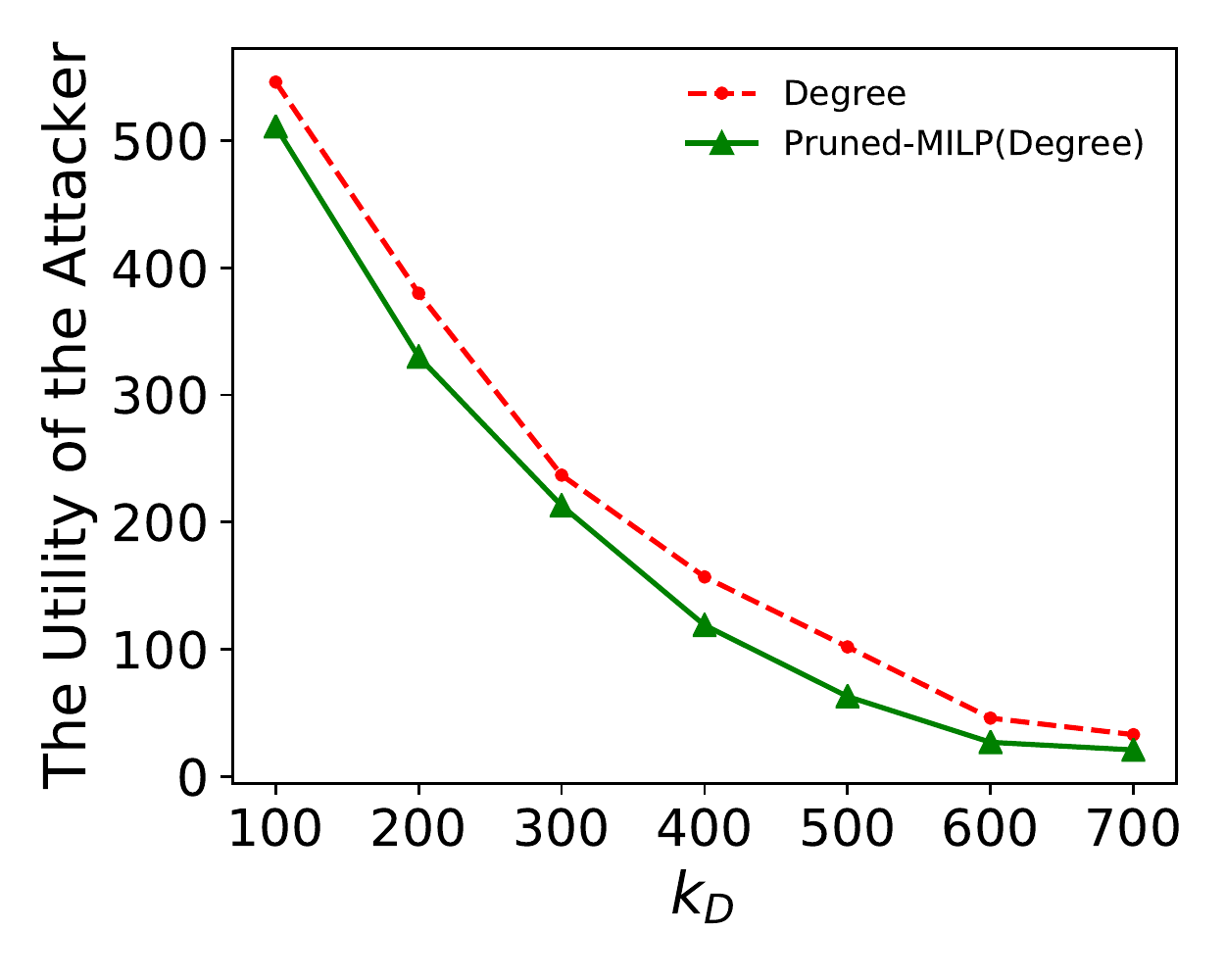}
	}%
	\caption{The performance of \texttt{PRUNED-MILP} on Hamsterster and Email-Eu-core networks against \texttt{k-MaxVD}.} 
	\label{kmaxVD}
\end{figure}

\begin{figure*}[th!] 
	\normalsize  
	\centering
	\subfigure[Email-Eu-core]{
		\includegraphics[scale=0.185]{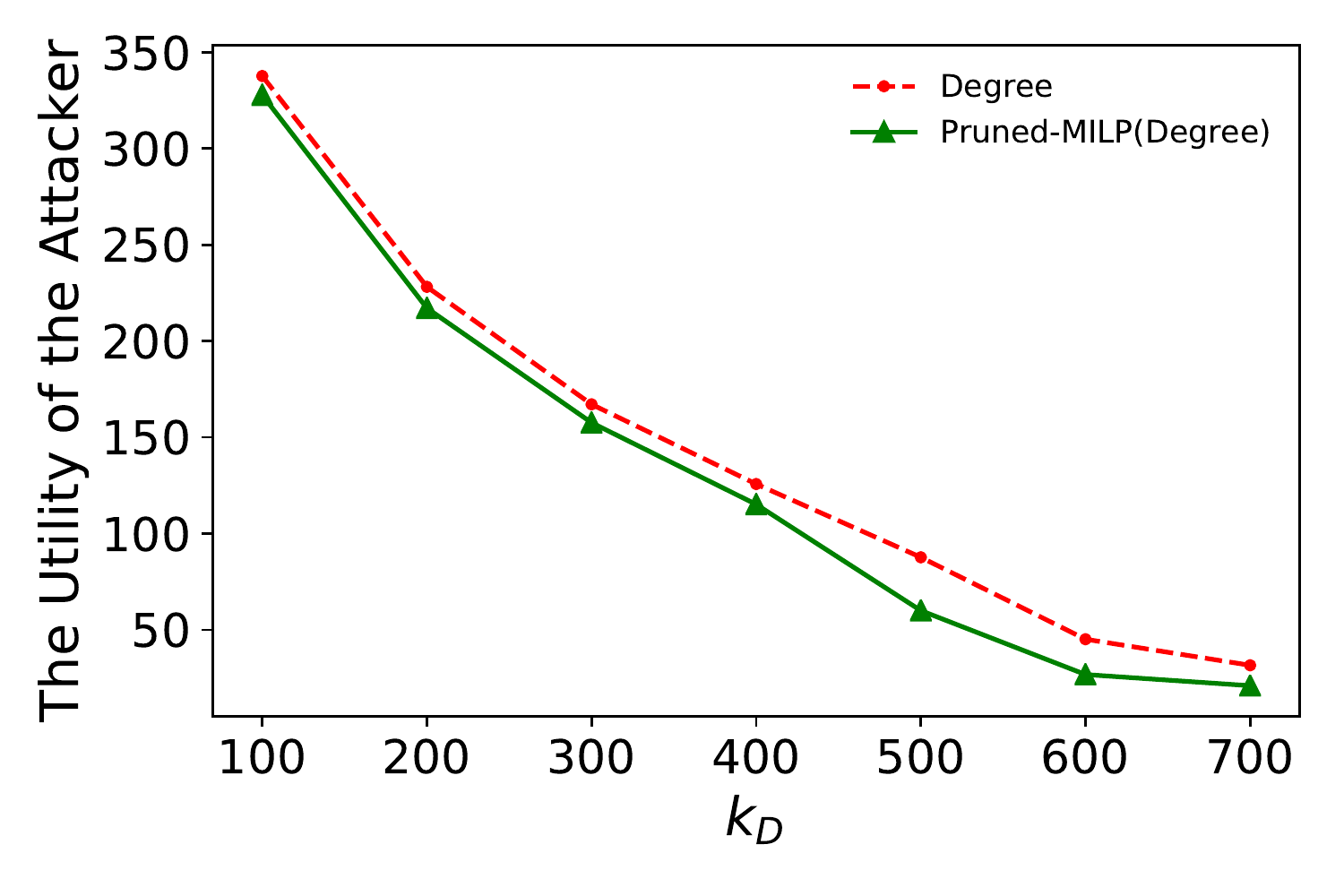}
	}%
	\subfigure[Hamsterster,$0.03$]{
		\includegraphics[scale=0.185]{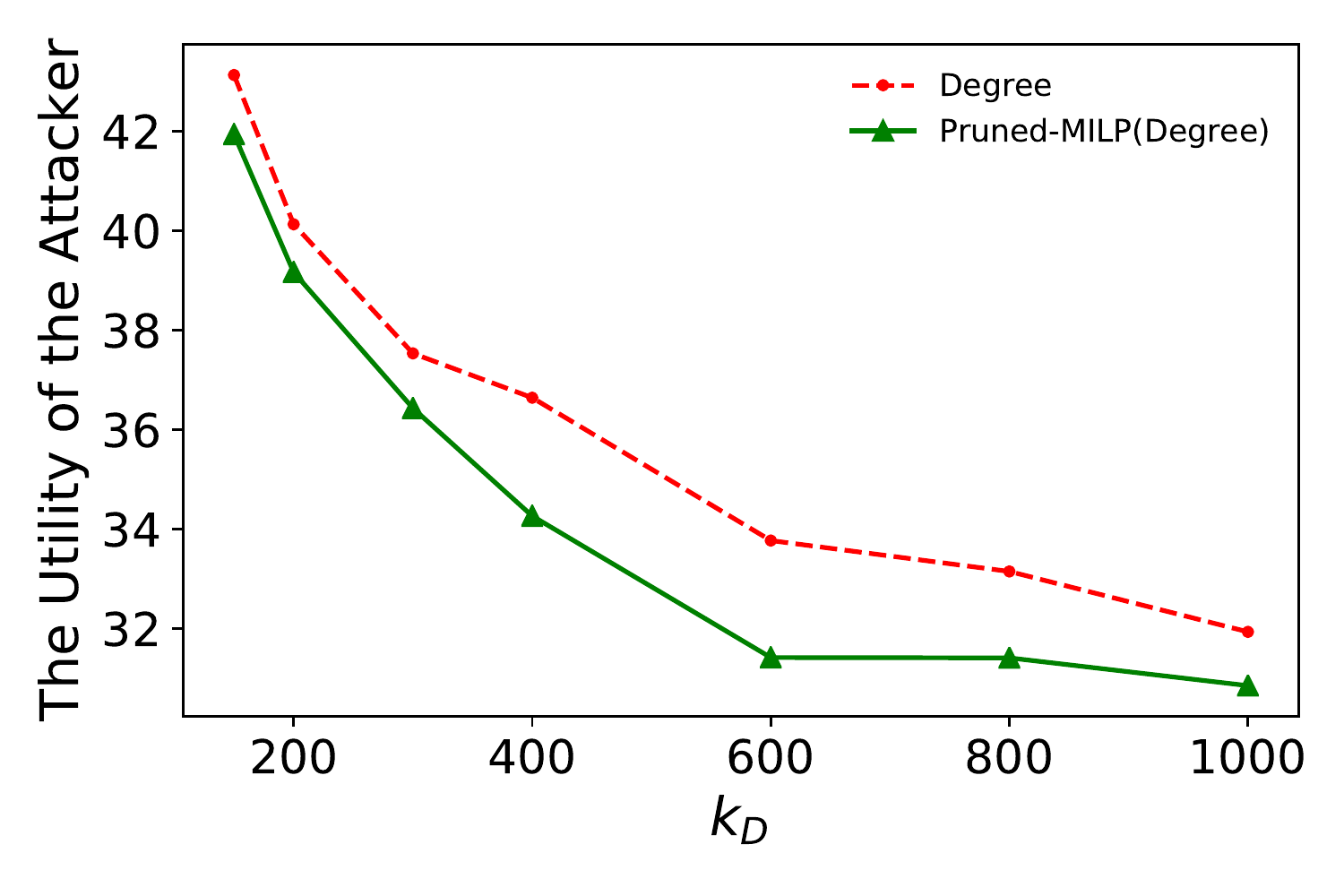}
	}%
	\subfigure[Hamsterster,$0.4$]{
		\includegraphics[scale=0.185]{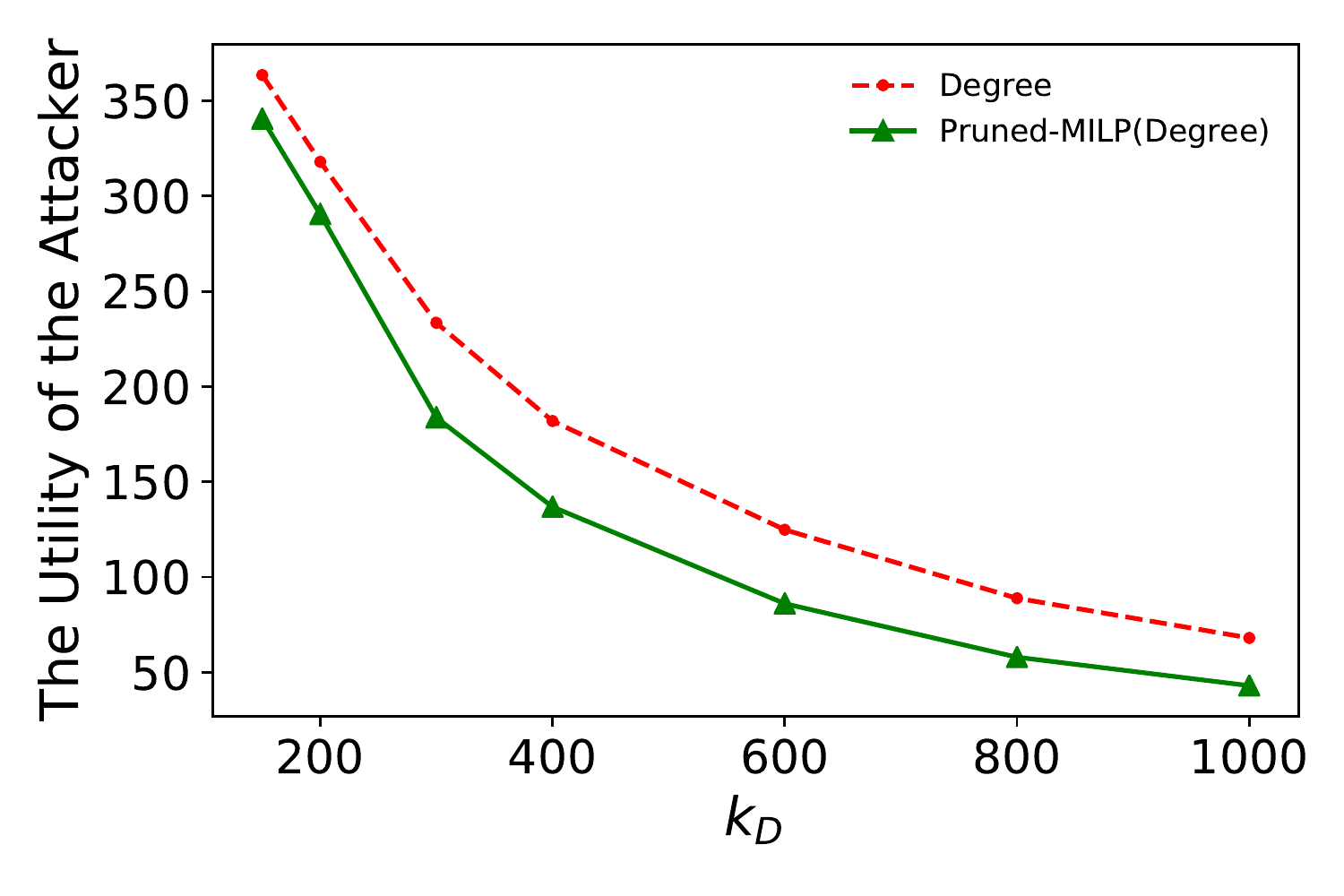}
	}%
	\subfigure[FB606(IC)]{
		\includegraphics[scale=0.185]{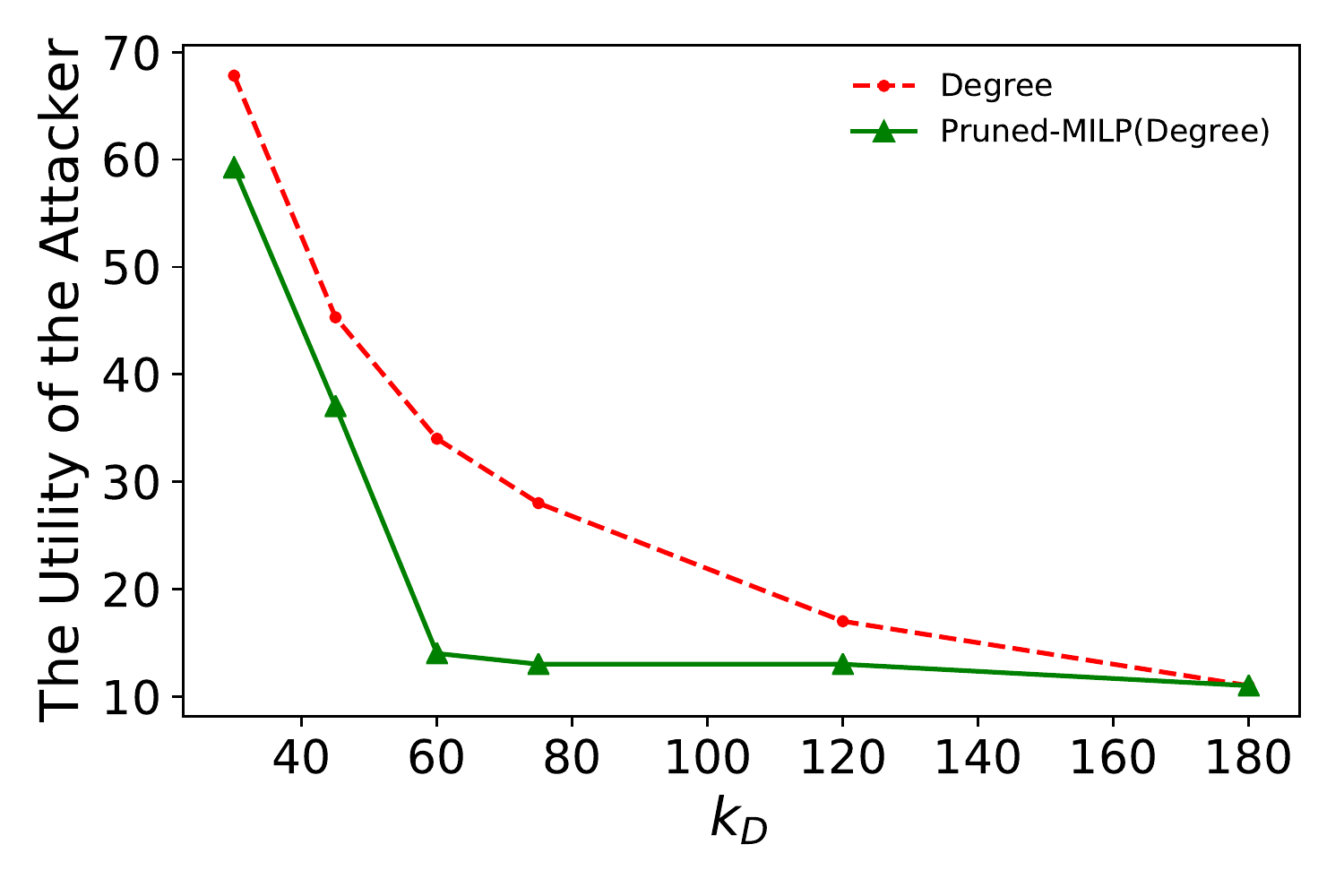}
	}%
	
	\subfigure[Enron3600,$0.03$]{
		\includegraphics[scale=0.185]{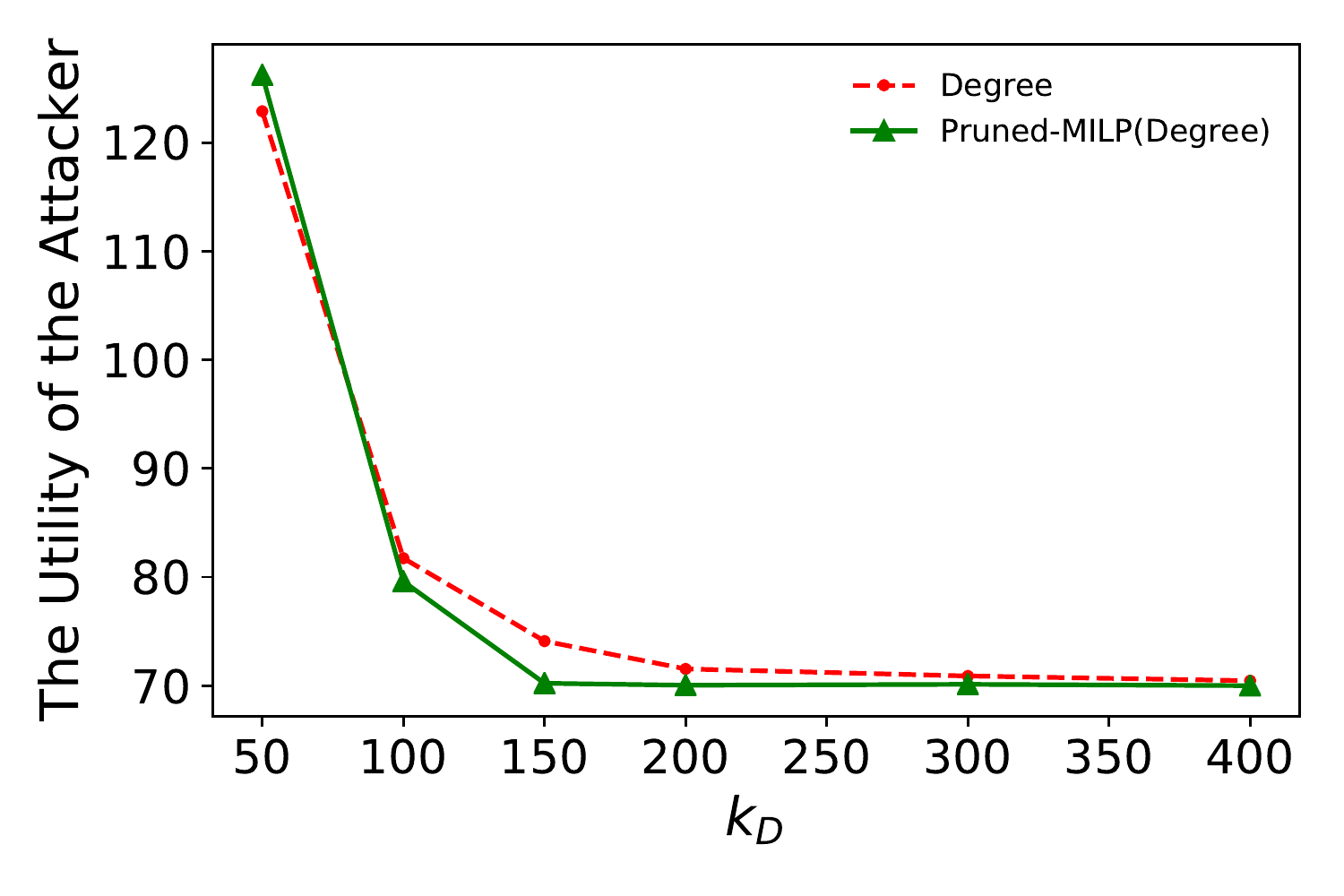}
	}%
	\subfigure[Enron3600,$0.4$]{
		\includegraphics[scale=0.185]{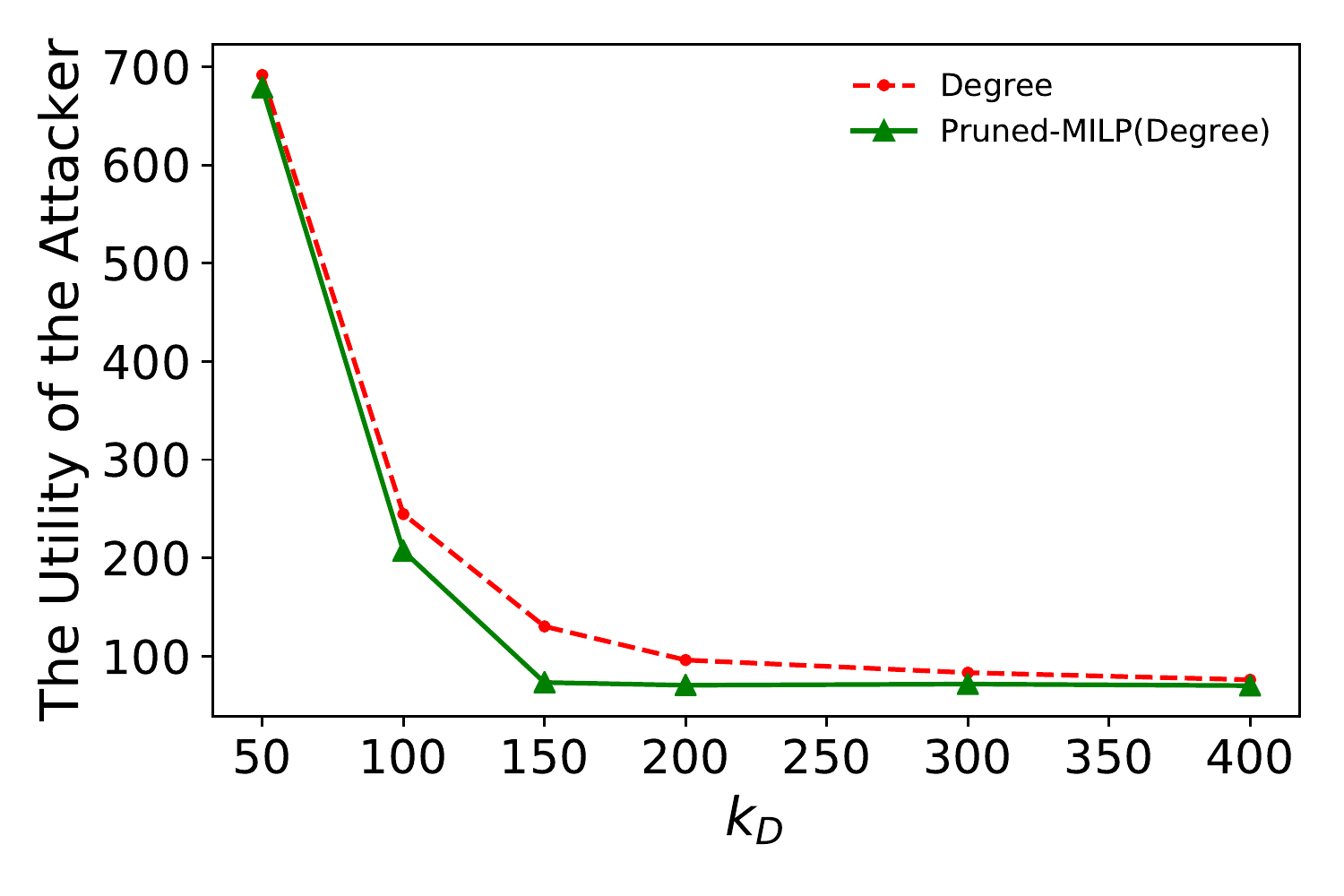}
	}%
	\subfigure[Enron4300,$0.03$]{
		\includegraphics[scale=0.185]{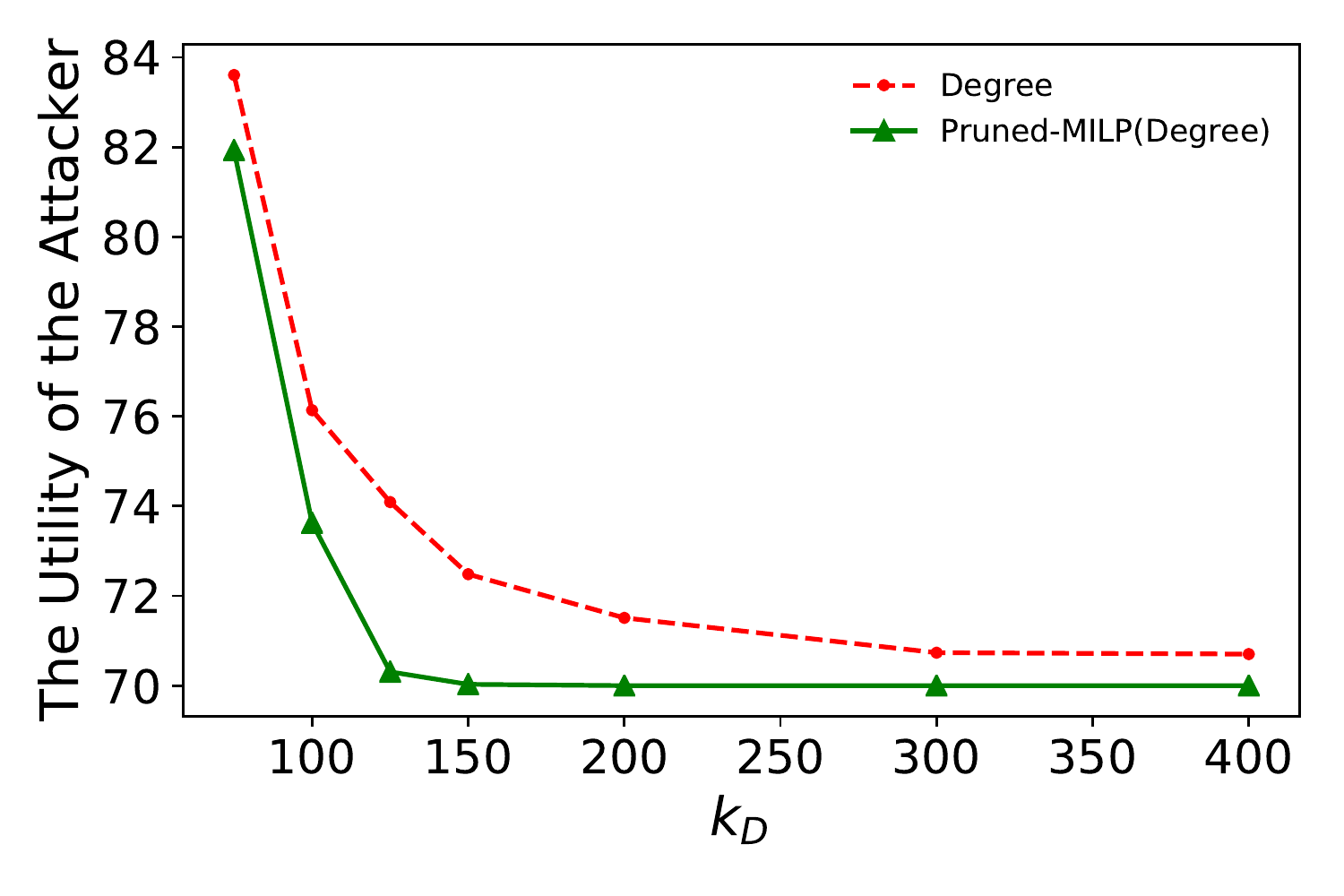}
	}%
	\subfigure[Enron4300,$0.4$]{
		\includegraphics[scale=0.185]{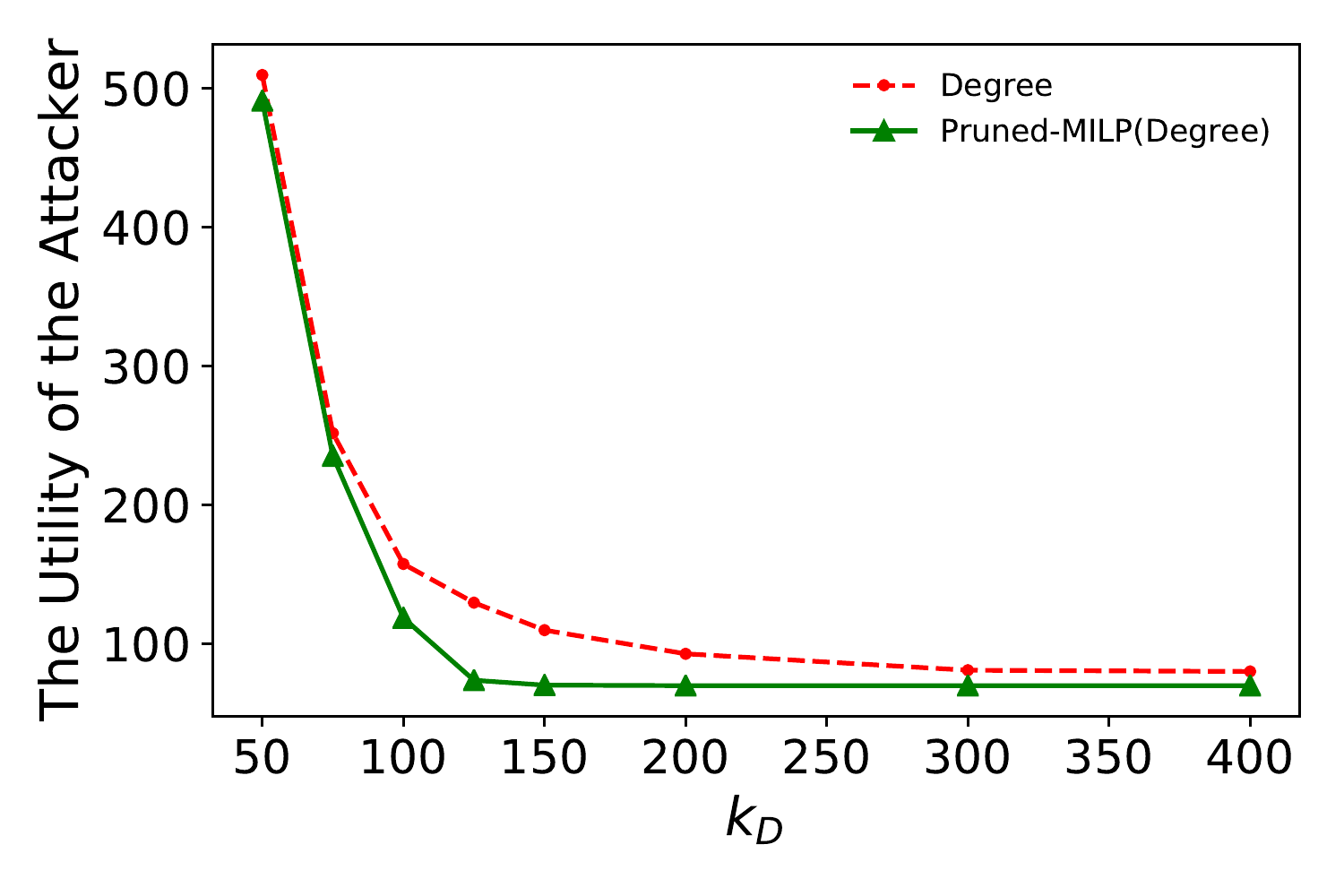}
	}%
	
	\subfigure[FB2000,$0.01$]{
		\includegraphics[scale=0.185]{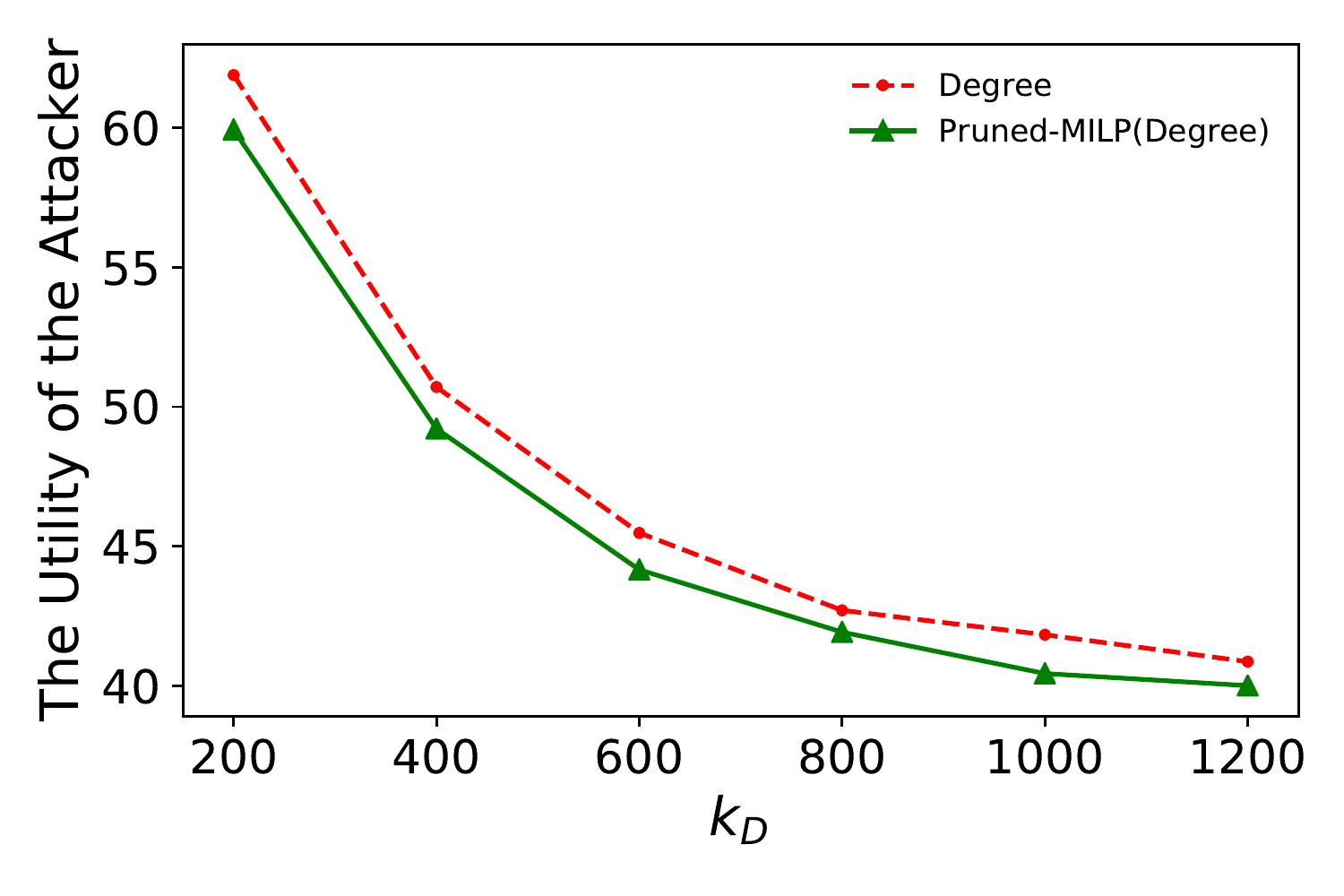}
	}
	\subfigure[FB3000,$0.01$]{
		\includegraphics[scale=0.185]{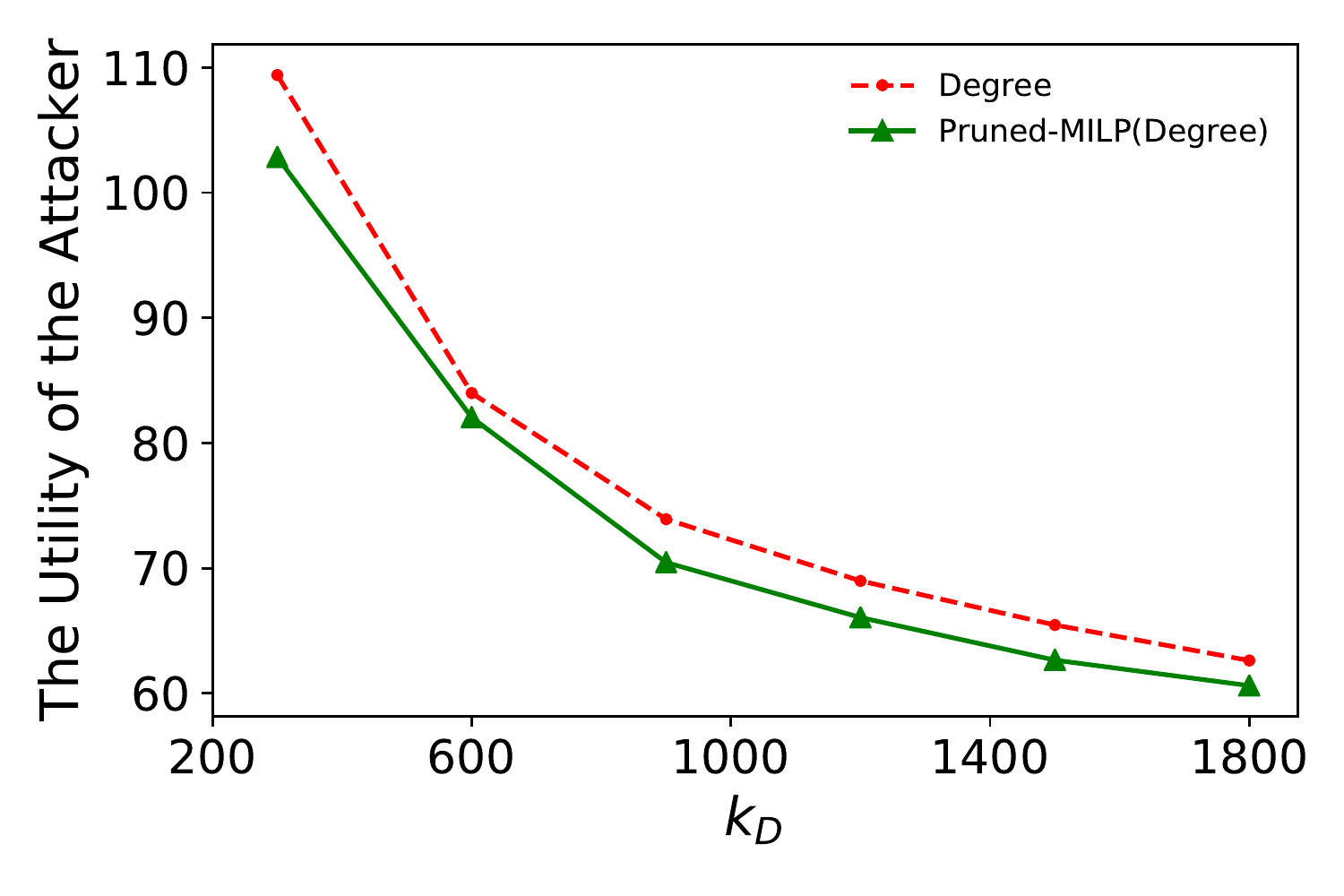}
	}
	\subfigure[FB2000,$0.1$]{
		\includegraphics[scale=0.185]{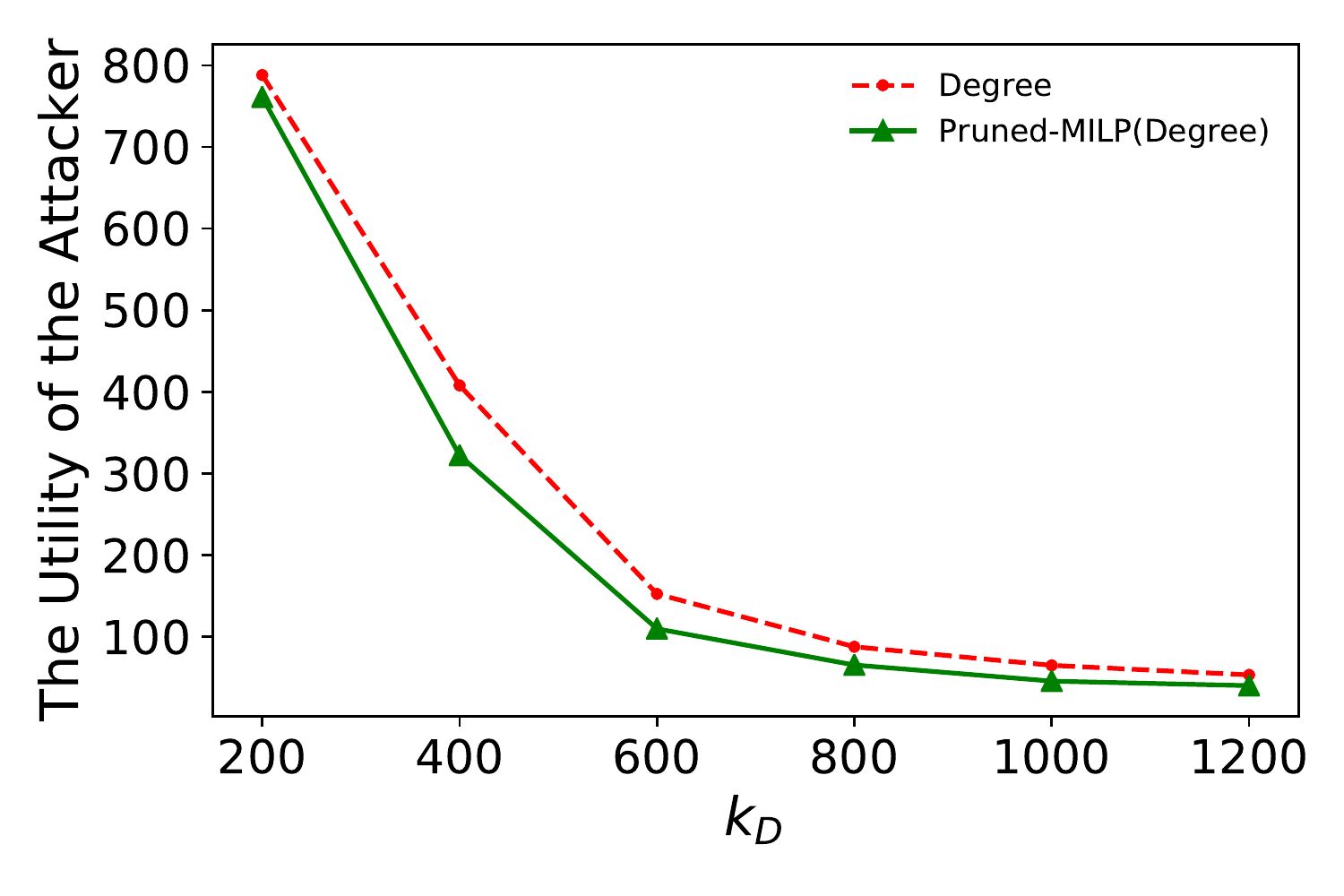}
	}
	\subfigure[FB3000,$0.1$]{
		\includegraphics[scale=0.185]{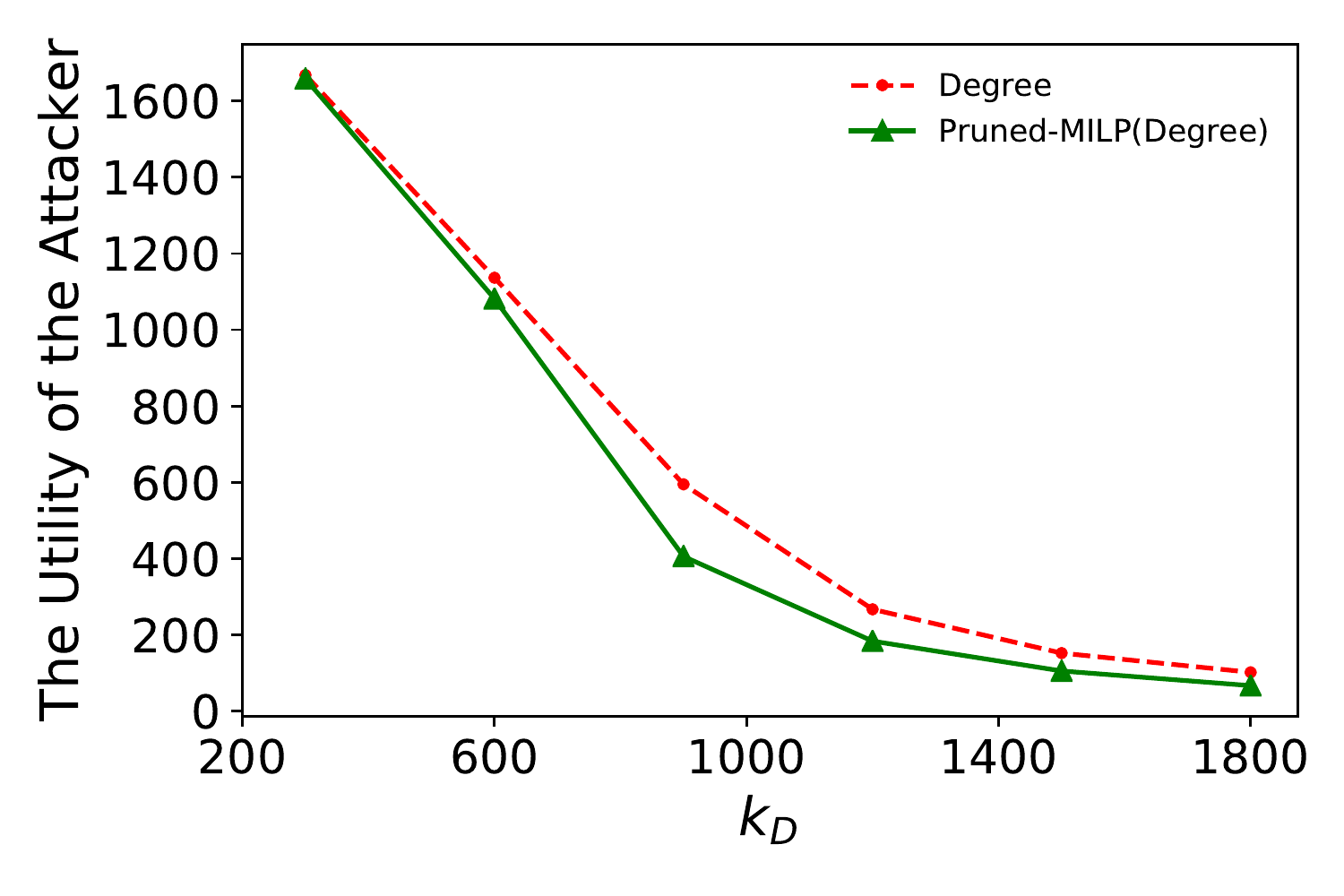}
	}
	\caption{The performance of \texttt{PRUNED-MILP} on real-world networks against \texttt{IM} attackers.} 
	\label{InfMax}
\end{figure*}

\begin{figure*}[!t]  
	\centering
		\includegraphics[scale=0.2]{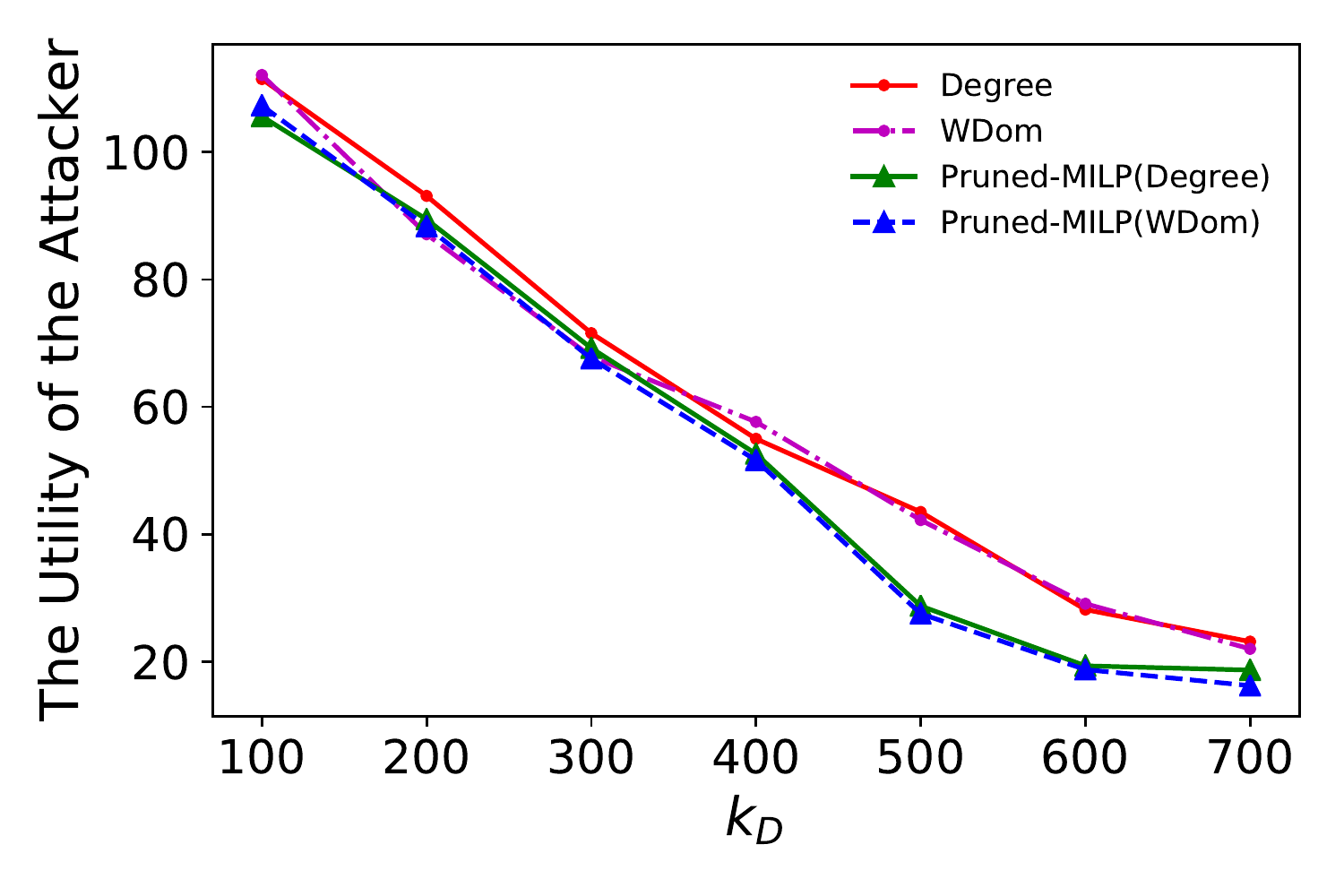}
		\includegraphics[scale=0.2]{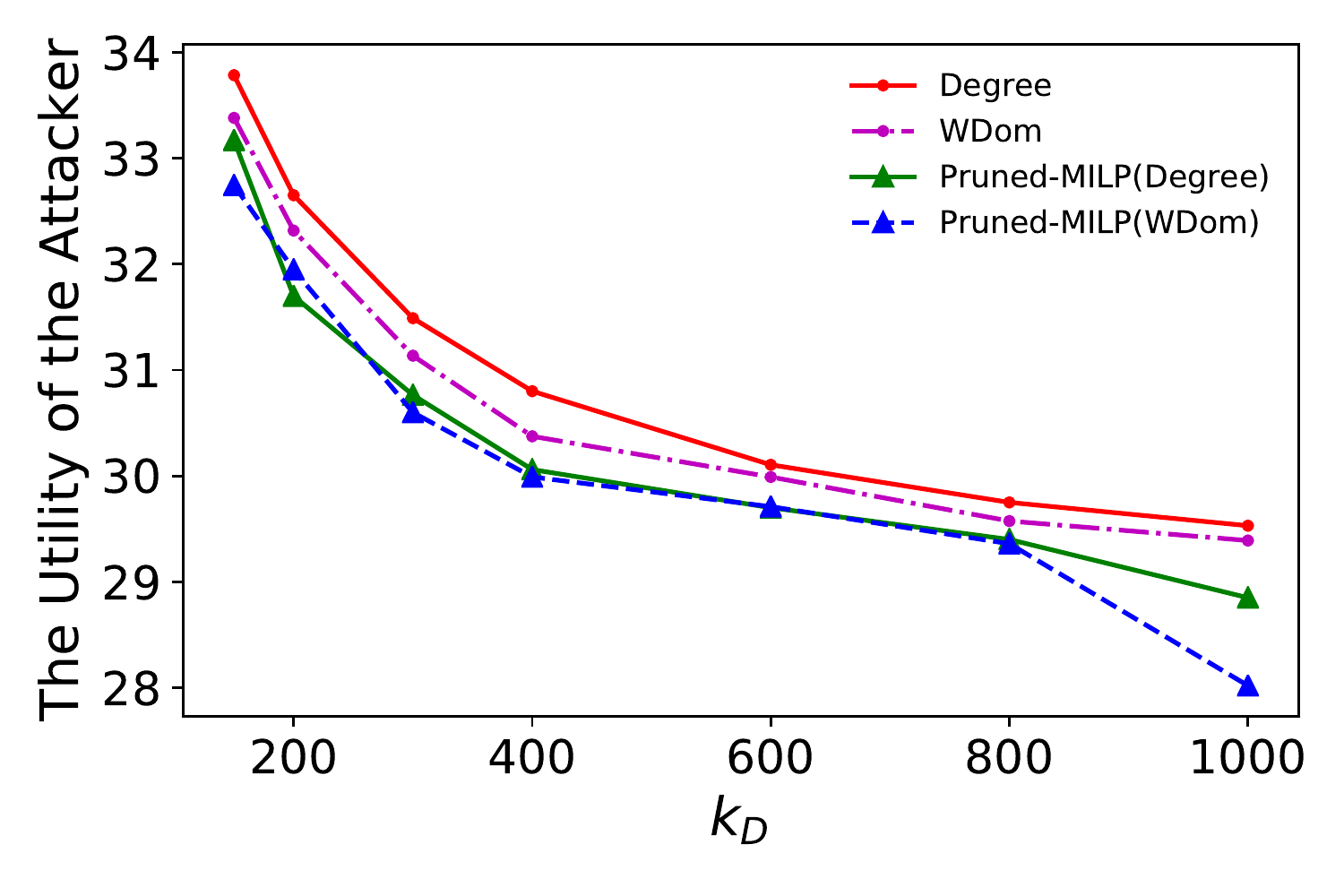}
		\includegraphics[scale=0.2]{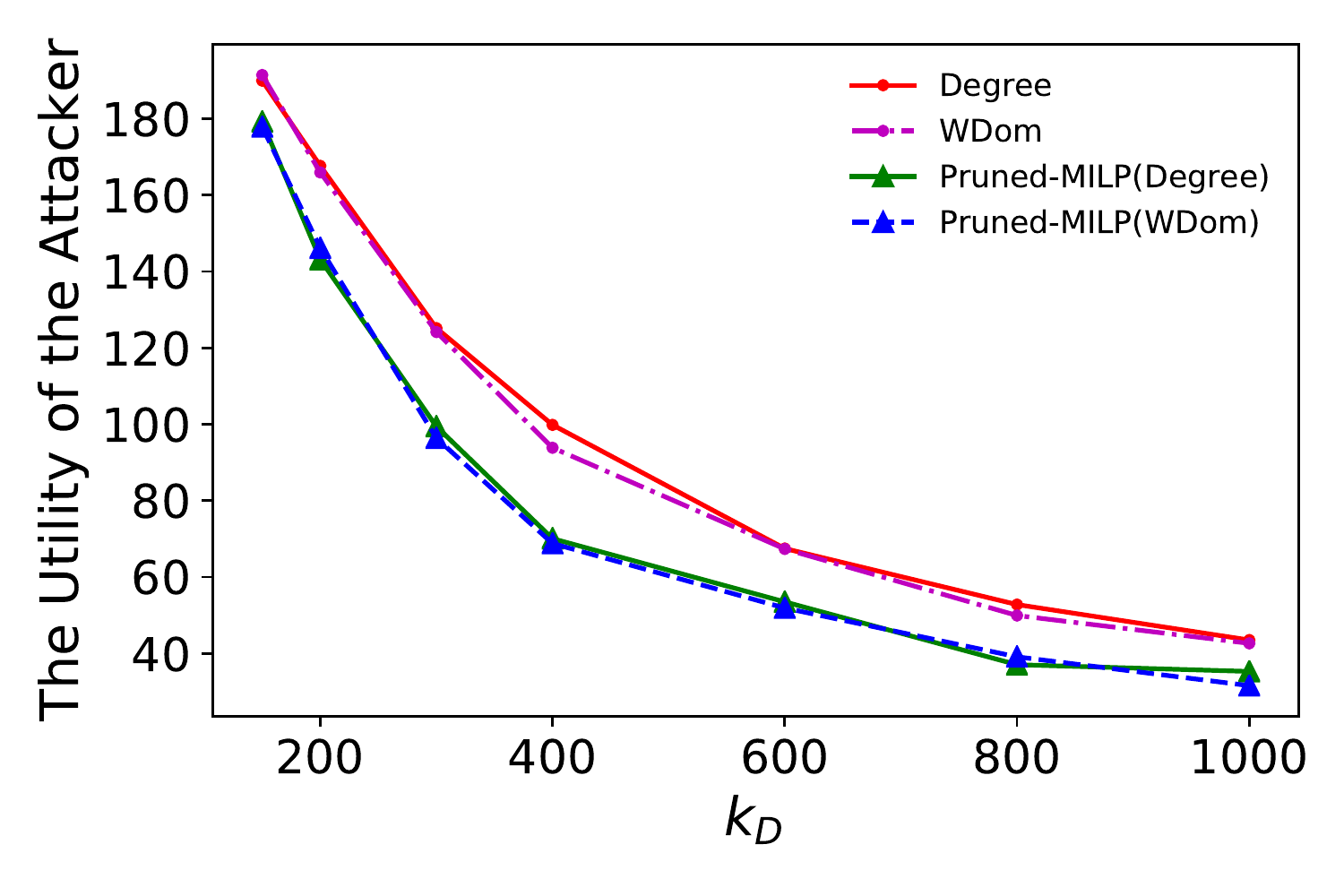}
	\caption{The performance of \texttt{PRUNED-MILP} on real-world networks against \texttt{WIM} attackers. Left: Email-Eu-core, $k_A = 20$. Middle: Hamsterster, $k_A = 30$, $p = 0.03$. Right: Hamsterster, $k_A = 30$, $p=0.4$.} 
	\label{wdom}
\end{figure*}

\paragraph{Defenses} 
Our optimal defense strategy is constraint generation(CG), and the primary defense strategy is DEF-MILP, where the defender solves the MILP \eqref{final} to find the set of nodes to block. We also consider DEF-WMILP, which is a variation of DEF-MILP in the weighted setting, as well as the corresponding pruning algorithms.

We compare our defense strategies with the following baselines. First, we consider a class of heuristic defense approaches where the defender blocks nodes in descending order of a specific node centrality measurement. 
The intuition is that node centrality measures the importance of a node in the network and blocking nodes with high centrality is more likely to limit the influence. In the experiment, we use node degree (out-degree in case of directed graphs), betweenness, PageRank, and influence as the centrality measurements and term the corresponding defenses as \texttt{Degree}, \texttt{Betweenness}, \texttt{PageRank}, and \texttt{Influence}, respectively. 
Specifically, the influence of a node is measured by the number of influenced nodes in the network when it is treated as the sole seed. 
We also consider four other baselines: \texttt{Influence Maximization (IM)}, \texttt{Greedy}, \texttt{WDom}, and  \texttt{Random}. In \texttt{IM}, the defender acts as an influence maximizer and blocks $k_D$ nodes that would cause the maximum influence. \texttt{Greedy} is a heuristic approach proposed in \cite{Wang2013NegativeIM}. They assume that an attacker chooses some influential nodes at the beginning, and a protector blocks the nodes according to the maximum marginal gain rule. In our experiment, we set the influential nodes as the seeds selected by influence maximization in the original network. In \texttt{WDom}, we define a quantity $\text{WDom}_j = \sum_{v_i\in N^I(v_j)}\mu_i$ for a node $j$, where $\mu_i$ is the non-negative weight of the node $i$, as the sum of weights of node $j$'s dominating nodes. This heuristic is used to defend \texttt{WIM} attackers by blocking $k_D$ nodes with highest $\text{WDom}$.
Finally, \texttt{Random} selects a random set of nodes to block.

\paragraph{Comparison with constraint generation}
First, we compare DEF-MILP with the constraint generation (CG)
algorithm.
We consider several variations of CG using a \emph{gap} parameter,
which defines the gap between solution quality of newly generated
constraint (i.e., attack) and the best previously generated
constraint; a gap of 0 implies that CG computes an optimal solution,
whereas other gaps trade off optimality and efficiency.
We evaluate the algorithms on ER networks whose sizes increase from $N = 15$ to $N = 65$. For each network's size, we generate 50 instances to test the runtime and 25 instances to test the attacker's utility with various random seeds and take the average. 
The experiments are conducted under DEF-MILP defense and CG defenses with budget $k_D = 5$,against the \texttt{K-MaxVD} attack with $k_A = 5$.  

We can see that the results of DEF-MILP are quite close to that of
optimal CG solutions, and are in some cases better than CG that uses a small gap.
Though the DEF-MILP is not far from the optimal solution, the runtime
is significantly reduced.
Fig.~\ref{fig:cg} shows that even if we loosen the gap of CG
algorithms to 1, 2, and 3, the runtime of DEF-MILP is still
considerably lower.

\paragraph{Results on synthetic graphs}

In our experiments, we generate $64$-node graphs for \texttt{k-MaxVD} and \texttt{IM} (including IC and LT) attackers with budgets $k_A = 5$. We generate $80$-node networks for \texttt{WIM} attackers with $k_A = 6$. Each node is associated with a value $\mu_i \sim U[0,1]$.

As shown in Fig.~\ref{random}, our defense strategy DEF-MILP and DEF-WMILP outperform all other baselines under all three attacks. 
We note that on BA graphs, heuristics based on node importance is comparable to the MILPs, while all these approaches perform significantly better than \texttt{Random}. One possible reason is that in BA graphs, there are a few high-degree nodes that can be effectively identified by centrality based algorithms.





For the defense algorithms, we can see that several heuristics can work effectively. \texttt{RageRank} is a good heuristic under \texttt{k-MaxVD} attack. \texttt{IM} works better than other heuristics under the LT model in our experiments.
For the \texttt{WIM} attacker, \texttt{WDom} heuristic can be slightly better than other heuristics.
 
 
\paragraph{Results on real-world networks}


As the size of real-world networks is significantly larger, we only test \Call{Pruned-MILP}{} and prune the nodes in descending order of the degrees. We compare \Call{Pruned-MILP}{} with \texttt{Degree} that uses the same node property to select nodes. 

Fig.~\ref{kmaxVD} shows the utility of the \texttt{k-MaxVD} attacker
on Hamsterster friendships network, with $k_A = 30$, and Email-Eu-core
network, with $k_A = 20$. The results show that our proposed approach
outperforms the Degree algorithm, even though aggressive pruning is used.

Fig.~\ref{InfMax} shows the defense of the \texttt{IM} attackers with different diffusion models in three networks. Linear Threshold (LT) model is used in the Email-Eu-core network. 
Uniform Independent Cascade (UIC) with different propagation probabilities are used in Hamsterster friendship network and the Weighted Independent Cascade (IC), in which each edge from node $u$ to $v$ has the propagation probability $1/deg(v)$ to activate $v$, is used in a 606-node sampled Facebook network. The budgets of the attacker are set as $k_A = 20$, $k_A = 30$ and $k_A = 10$, respectively.

\begin{table}[ht]
\centering
\begin{tabular}{@{}l|lll|lll@{}}
\toprule
\multirow{2}{*}{$k_A$} & \multicolumn{3}{l|}{Hamsterster} & \multicolumn{3}{l}{Email-Eu-core} \\ \cmidrule(){2-7} 
                       & $M_{LP}$         & $M_{MILP}$  & Gap(\textperthousand )    & $M_{LP}$         & $M_{MILP}$     & Gap(\textperthousand )     \\ \midrule
10                     & 320.500      & 320.000    & 1.563       & 690.380     & 689.000     & 2.002        \\
20                     & 443.000    & 443.000    & 0.000          & 784.000        & 782.000     & 2.558        \\
30                     & 531.875    & 531.000    & 1.648       & 836.500      & 836.000     & 0.598        \\
40                     & 603.750     & 603.000    & 1.244       & 872.090     & 872.000     & 0.103        \\
50                     & 660.500      & 660.000    & 0.758      & 895.830     & 895.000     & 0.927        \\
60                     & 707.375    & 707.000    & 0.530      & 915.839     & 915.000     & 0.917        \\ \bottomrule
\end{tabular}
\caption{The effect of integrality relaxation of BR-MILP (\ref{Eqn-MILP}) in Hamsterster and Email-Eu-core networks.} 
\label{gap}
\end{table}


For the larger two network datasets, we evaluate the performance of \Call{PRUNED-MILP}{} in their sub-networks. Enron email network is sampled to Enron3600 containing 3,600 nodes and 11,412 edges and Enron4300 containing 4,300 nodes and 11,968 edges ($k_A = 70$). Facebook network is sampled to sub-networks with 
2000 ($k_A = 40$) and 3000 ($k_A = 60$) 
nodes. The two sets of networks are applying different UIC model in view of small ($p = 0.01$, $p = 0.03$) and large ($p = 0.1$, $p = 0.4$) diffusion probabilities. 
Fig.~\ref{InfMax} shows that our algorithm is generally better than the Degree algorithm. 



Next, we evaluate our \Call{PRUNED-MILP}{} defense of \texttt{WIM} attack.
Fig.~\ref{wdom} shows the utility of the attacker on the Email-Eu-core network and the Hamsterster friendship network. Each node $v_i\in V$ in the networks is assigned a value $\mu_i$ uniformed distributed in [0,1].

We compare the two pairs of experiments with two kinds of pruning orders, \texttt{Degree} and \texttt{WDom}. Intuitively, \texttt{WDom} considers the value of nodes so that it might be more adaptable to this problem. Fig.~\ref{wdom} shows that applying the proposed \Call{PRUNED-MILP}{} outperforms the original defense strategies. 


\paragraph{The effect of LP relaxation}

In our approach, we relaxed the integral constraints on the variables $\mathbf{y}$ of the BR-MILP, through which we are essentially optimizing over an upper bound of the attacker's utility.
We demonstrate the quality of this approximation through experiments. Let the relaxed problem be BR-LP. We compare the optimal objective values of BR-MILP and BR-LP, denoted as $M_{LP}$ and $M_{MILP}$, respectively.
We are interested in the integrality gap defined as  $IG = M_{LP}/M_{MILP}$. Table \ref{gap} shows the gap in percentage, defined as  $\text{Gap} = (M_{LP}-M_{MILP})/M_{MILP}$, for the Hamsterster network and Email-Eu-core network with the attackers' budget from 10 to 60. The results show that the gaps in various cases are almost negligible, demonstrating a good approximation quality at least from an experimental perspective.
The experiments in synethic networks achieves similar results.

\paragraph{Trade-off in Heuristic Pruning Algorithm}

\begin{table}[htp]
\begin{center}
\begin{tabular}{@{}l|lllll@{}}
\toprule
$l_d$              & 400.0 &500  & 550.0   & 600.0    & 639.0   \\ \midrule
Run-time (sec)     & 7.4 &12.8  & 17.4 & 131.0     & 589.0   \\
$U_{IM}$ & 188.7 & 153.7 & 146.8 & 138.4  & 133.1 \\
$U_{k-\text{MaxVD}}$  & 210.0  & 194  & 175.0   & 164.0     & 155.0   \\ \bottomrule
\end{tabular}
\end{center}
\caption{The run-time and solution quality in Hamsterster network with $k_D = 400$ and $k_A = 30$} 
\label{tradeoff}
\end{table}
The parameter $l_d$ in our pruning algorithm trades off the run-time and quality of the algorithm. In Table \ref{tradeoff}, we show the run-time and the attacker's utilities with different configurations of $l_d$ in Hamsterster network when $k_D = 400$ and $k_A = 30$. 
$U_{k-\text{MaxVD}}$ denotes the utility of the \texttt{$k$-MaxVD} attacker, and $U_{IM}$ denotes the utility of the \texttt{IM} attacker with propagation probability $p = 0.4$ . 
We can observe that when $l_d$ increases, runtime quickly increases,
but the solution quality also improves. However, when $l_d$ is larger
than one threshold, CPLEX cannot return the solution in reasonable
time.